\documentclass[12pt]{article}

\usepackage{a4wide,hyperref}

\usepackage[T2A]{fontenc}
\usepackage[utf8]{inputenc}
\usepackage{graphicx}
\usepackage{amsmath,amssymb,amsthm,amsfonts,amscd}
\usepackage{float}
\makeatletter
\@ifundefined{IfDocumentMetadataTF}{}{}
\makeatother
\begin{document}
\date{}

\def\le{\leqslant}
\def\leq{\leqslant}
\def\ge{\geqslant}
\def\geq{\geqslant}
\def\epsilon{\varepsilon}
\def\phi{\varphi}
\def\hat{\widehat}
\def\tilde{\widetilde}
\def\bar{\overline}
\def\cal{\mathcal}
\def\kappa{\varkappa}
\def\frak{\mathfrak}

\def\grad{{\bf{grad}}}
\def\rot{{\operatorname{rot}}}
\def\div{{\operatorname{div}}}
\def\v{{\vec{\bf v}}}
\def\B{\vec{\bf B}}
\def\E{\vec{\bf E}}
\def\R{{\Bbb R}}
\def\A{{\bf A}}
\def\x{{\bf x}}
\def\y{{\bf y}}
\def\z{{\bf z}}
\def\V {{\vec{\bf V}}}
\def\Z{\Bbb{Z}}
\def\b{\vec{\bf b}}
\def\r{\vec{\bf r}}
\def\B{\vec{\bf B}}
\def\G{\vec{\bf G}}
\def\U{\bf U}
\def\u{\vec{\bf u}}
\def\F {{\vec{\bf F}}}
\def\i{{\bf i}}
\def\j{{\bf j}}
\def\r{{\bf r}}
\def\k{{\bf k}}
\def\H{{\bf H}}
\newcommand{\ee}{\varepsilon}
\newcommand{\pa}{\partial}

\newtheorem{theorem}{\bf Theorem}[section]
\newtheorem{proposition}{\bf Proposition}[section]
\newtheorem{lemma}{\bf Lemma}[section]
\newtheorem{corollary}{\bf Corollary}[section]

\theoremstyle{remark}
\newtheorem{remark}{\bf Remark}[section]
\newtheorem{definition}{\bf Definition}[section]
\newtheorem{example}{\bf Example}[section]
\numberwithin{equation}{section}
\sloppy

\setcounter{page}{1}

\title{\textbf{Magnetohydrodynamic spectra and line invariants in the presence of axions}}

\author{P.~M.~Akhmetiev$^{a}$;
	M.~S.~Dvornikov$^{a}$; 
	A.~Yu.~Smirnov$^{a,b}$ \\
	\small{$^a$Pushkov Institute of Terrestrial Magnetism, Ionosphere}
	\\
		\small{and Radiowave Propagation (IZMIRAN)}\\
	\small{$^b$Faculty of Physics, Lomonosov Moscow State University}}

\maketitle

\begin{abstract}
	A new form of the mean-field equations for magnetic fields, which accounts for the interaction with axions, is studied. It is shown that the asymptotic ergodic invariants of magnetic field lines play a key role in the evolution of magnetic fields, which determine the type of the magnetohydrodynamic spectrum of the solutions.
\end{abstract}

\section{Introduction}

The existence of dark matter, which our universe mainly consists of, is the observationally confirmed fact. The nature of dark matter constituents is unknown still. After the Large Hadron Collider significantly constrained the parameter space for supersymmetric particles (see, e.g., Ref.~\cite{Can19}), the most probable candidates for dark matter became light particles called axions or axion like particles (ALPs)~\cite{Luz20}.

We recall that axions were first suggested in the quantum chromodynamics (QCD) in Refs.~\cite{PecQui77,Wei78,Wil78} to solve the strong CP problem. The mass of a QCD axion is in the range $10^{-6}\,\text{eV}<m<10^{-4}\,\text{eV}$. The mass of ALP can be much smaller, $m<10^{-11}\,\text{eV}$~\cite{Mar16,Cho21}. Numerous attempts, undertaken to detect axions and ALPs in a laboratory~\cite{IraRed18,Sik21}, are unsuccessful still.

The only sizable interaction, which axions and ALPs participate in, is gravity. Nevertheless, these particles can slightly interact with electromagnetic fields~\cite{KimCar10}. This weak electromagnetic interaction can result in various phenomena like
the emission of ultra-high-energy cosmic rays, gamma ray bursts with certain energies, etc.~\cite{Car25}.

The axion production happens mainly in the early universe. In this situation the axionic background is spatially homogeneous.
Nevertheless, if the Peccei--Quinn (PQ) phase
transition happens after the reheating, axions can become inhomogeneous. In this situation, the formation of axion
miniclusters~\cite{KolTka93,KolTka94} and axion stars~\cite{ChaDel11,BraZha19} is possible.

The axion electrodynamics was formulated in Refs.~\cite{Sik83,Wil87}. However, if axions interact with large scale magnetic fields, one should deal with the axion magnetohydrodynamics (MHD), which was put forward in Ref.~\cite{LonVac15}, where it was shown that magnetic fields become unstable in presence of axions. The axionic MHD in the early universe was studied in Refs.~\cite{DvoSem20,Dvo22}.

Axions can be present in stars like the Sun. Solar axions are produced, e.g., in the Primakoff effect in an external magnetic field~\cite{Car25}. However, axionic dark matter can be accumulated in the Sun from the earlier epochs of the universe~\cite{DilZio03,Zhi17}. Solar axions can be searched with various types of helioscopes (see, e.g., Ref.~\cite{Car25} for the recent development).

In Refs.~\cite{DvoAkh24,AkhDvo24}, we developed axion MHD by proposing the new
induction equation for the magnetic field accounting for inhomogeneous axions. This equation was applied in Refs.~\cite{Dvo24,Dvo25} to study magnetized axionic clumps embedded in solar plasma. In Refs.~\cite{Dvo24,Dvo25}, we put forward an idea that the decay of these clumps can contribute to the solar corona heating. The axion dynamo in neutron stars was studied in Ref.~\cite{Anz23}.

Nowadays it is believed (see, e.g., Ref.~\cite{Sti04}) that the solar magnetic activity is likely to be driven by the Parker's dynamo mechanism~\cite{Par55}. Nevertheless, this kind of dynamo is not a unique one. We suggest that axionic dynamo, studied in Refs.~\cite{DvoAkh24,AkhDvo24,Dvo24,Dvo25}, can become an alternative for the Parker mechanism in some situations. In the present work, we continue our studies of axionic MHD and its applications to solar physics.

In this paper, we adopted an approach based on mean magnetic field theory \cite{Mo}. The problem is of interest since in the mean field equations, which traditionally have only two terms (the $\alpha$-term associated with the field generation and the $\beta$-term associated with the polar dissipation), one adds a third $\gamma$-term and also takes into account the axion wave equation, which is discussed shortly in Sec.~\ref{appl} we
briefly recall calculations from \cite{AkhDvo24},\cite{A-D2} and apply the balance relationships (\ref{pokaz}), (\ref{pokaz2}) for dynamo between the terms in this system of equations. To collect all results, the magnetic field interacting with axions can exist in two fundamental states. They are (i) a
configuration with a low magnetic energy and a zero helicity, which we called the axion mirror soliton and associated with the Kraichnan MHD spectrum;  (ii) a configuration with increasing magnetic modes under the action of an axion dynamo, which we associated with the Kolmogorov MHD spectrum. The names of the MHD spectrum types were defined in Ref.~\cite{Z-R-S}.

The key role in solving the axion mean field equation is played by the higher invariants of magnetic lines in the ergodic asymptotic form proposed in Ref.~\cite{Arn1}.
We use the $M_3$ invariant, whose integral representation for a field with closed magnetic lines is available in Ref.~\cite{A2}. It is widely believed that no magnetic line invariants other than the Gauss coefficients of magnetic line links exist.

The reason for this fact was discovered in Ref.~\cite{K}. It consists in the fact that all higher magnetic line invariants are unstable and are destroyed by small perturbations of the magnetic field. This instability phenomenon requires further study. We believe that such instability does not manifest itself in random mean field theory. It means that the M-invariant is ergodic.

\section{Mean field equation and MHD}\label{MHD}

The system of MHD equations (see, for example, Ref.~\cite[Ch.~V, Sec.~1, Remark~1.6]{A-Kh} ) for an ideal approximation (i.e., for an incompressible, inviscid, and an infinitely conductive fluid without magnetic dissipation)
is nonlinear. The Kolmogorov's approach to hydrodynamics is linked to the probability theory and allows us to identify a specific spectrum of solutions based on physical considerations. A probabilistic approach to the MHD equations was also proven to be possible. It is called the mean field theory~\cite{Mo,Z-R-S}. Let us recall its basics.

We denote the hydrodynamic velocity field by $\V(\x,t)$, and the magnetic field in a liquid conducting medium by $\B(\x;t)$. We assume that the medium is homogeneous and isotropic. For simplicity, we also assume that the 3D region coincides with the entire space. Then the Fourier spectra for $\V$ and $\B$ are continuous.

We assume that the vector functions $\V$ and $\B$ are random and divide them into an average part, a large scale component, and a fluctuating component a small scale or turbulent component. The characteristic length of the small scale component, which is, generally speaking, not the component itself, is considered to be much smaller than the characteristic scale of measurement of the averaged quantities (for more details, see Ref.~\cite[Ch.~7, Eqs. (7.5) and~(7.6)]{Mo}):
\begin{eqnarray}\label{mean}
	\begin{array}{cc}
		\V(\x,t) = \V_0(\x,t) + \v (\x,t), \quad \langle \v \rangle = 0 \\
		\B(\x,t) = \B_0(\x,t) + \b (\x,t), \quad \langle \b \rangle = 0,
	\end{array}
\end{eqnarray}
where $\langle \v \rangle$ and $\langle \b \rangle$ denote the average fluctuation components.
This yields $\langle \V \rangle = \V_0$ and $\langle \B \rangle = \B_0$. A more detailed study assumes that a random ensemble is defined, and the averaging of the observed quantities is described by a limit transition for characteristic values of length and time.

Let us consider a two scale MHD model. 
Our goal is to investigate how the small scale component of the observed field
affects the magnetic helicity flux of the large scale component of the magnetic field.
The random fields
$\V$ and $\v$ are assumed to be given, whereas the vectors $\B_0$ and $\b$ should be calculated.
The key feature is the linearity of the induction equation in the magnetic field.

Let us separate the induction equation,
\begin{eqnarray}\label{induction}
	\frac{\partial \B(\x,t)}{\partial t} = \rot(\V(\x) \times \B(\x)) + \eta_{\rm{c}} \Delta(\B(\x))
\end{eqnarray}
into two evolution equations for the regular large scale part and for the fluctuating small scale part (see Ref.~\cite[Eqs~(7.7) and~(7.8)]{Mo}):
\begin{eqnarray}\label{fluctuation}
	\begin{array}{cc}
		\frac{\partial \B_0}{\partial t} = \rot( \V_0 \times \B_0)  + \rot(\Upsilon) + \eta_{\rm{c}} \Delta \B_0, \\
		\frac{\partial \b}{\partial t} = \rot( \V_0 \times \b) + \rot(\v \times \B_0) + \rot \G + \eta_{\rm{c}} \Delta \b,
	\end{array}
\end{eqnarray}
where $\Upsilon = \langle \v \times \b \rangle$ and $\G = \v \times \b - \langle \v \times \b \rangle$.  In Eq.~\eqref{fluctuation}, the coefficient $\eta_{\rm{c}}$ is scale-independent. It is called the magnetic diffusion coefficient, $\eta_{\rm{c}}=\sigma^{-1}_{\rm{c}}$. We remind the this coefficient is assumed to be infinitely small. In this case, the induction equation is said to be studied in the ideal approximation.
We examine the first of the equations for the regular part
$\B_0$ of the magnetic field $\B$ .This equation calculates the electromotive force $\Upsilon$ due to the fluctuating part.

Since the second equation provides a linear relation between
$\b$ and $\B_0$, the field
$\Upsilon$ and the field $\B_0$ are also related by a linear relation.
Assuming that the relation between
$\B_0$ and $\b$ is isotropic, the following two leading terms arise:
\begin{eqnarray}\label{alpha}
	\Upsilon = \alpha \B_0 - \beta \rot(\B_0).
\end{eqnarray}
The terms correspond to the decomposition of the random fields
$\B_0$ and $\b$ into symmetric and skew-symmetric parts. Since the coefficient $\beta$, which depends on the averaging scale, takes on the role of a diffusion coefficient, it is called the turbulent diffusion coefficient.
We assume that
\begin{eqnarray}\label{beta}
	\beta \sim \eta k^{a},
\end{eqnarray}
where $a \le 0$ is called the index. 

Under the assumption made in Eq.~(\ref{alpha}), we take the scalar product of both sides of the first equation in Eq.~(\ref{fluctuation}) with the vector $\B_0$. Additionally, we assume that $\eta_{\rm{c}}$ is very big, that means that molecular magnetic diffusion is not taken into account,
and take the integrals of both sides over the volume
$\Omega$, using the equality for $\A_0$, $\rot(\A_0)=\B_0$.
We obtain the following equation, which describes the transport of magnetic helicity in a large scale magnetic field
$\chi_{\B_0}= \int (\A_0,\B_0) d\Omega$:
\begin{eqnarray}\label{helicity4}
	\frac{d\chi_{\B_0}}{dt} = 2\alpha \int (\B_0,\B_0)d\Omega - 2\beta \int (\B_0,\rot(\B_0)) d\Omega.
\end{eqnarray}	

One of the integrals on the right-hand side of the equation has already been discussed:
$U_{\B_0}=2\int(\B_0,\B_0)d\Omega$. It is the magnetic energy of the average magnetic field.
The integral
$\chi_{\rot \B_0}=2\int (\B_0,\rot(\B_0)) d\Omega$ is called the current helicity integral
(of the large scale magnetic field).
The coefficients
$\alpha$ and $\beta$ play a fundamental role in MHD. They are not dimensionless.
The scaling exponents of these coefficients determine the properties of the solution.

We calculate the spectral indexes of the solutions using Kolmogorov's approach, which examines the Navier-Stokes equation. The MHD system in the absence of a magnetic field transforms into the Navier-Stokes equations. According to this theory and assuming homogeneity and isotropy, it turns out that the spectral indexes of the square of the hydrodynamic velocity in the steady state decrease with increasing wave numbers as
\begin{eqnarray}\label{5/3}
	(U(k),U(k)) \sim k^{-\frac{5}{3}},
\end{eqnarray}
where we take that  the hydrodynamic energy collected over all wave vectors of length $k$, both left and right.

A hypothesis
for the root-mean-square spacing
$\u^2(\vert\vert \delta \r\vert\vert)$
of the velocity $\U + \u(\delta \r)$ in a hydrodynamic flow:
\begin{eqnarray}\label{2/3}
	\u^2(\delta\r) \sim \vert\vert\delta\r\vert\vert^{\frac{2}{3}}.
\end{eqnarray}
This relation is called the Kolmogorov's $\frac{2}{3}$ law.

However, other spectra that can arise in a given problem are of interest besides the Kolmogorov one. For example, the well-known Kraichnan spectrum with the magnetic and hydrodynamic energy index $k^{-\frac{3}{2}}$.   For this spectrum one has that
\begin{eqnarray}\label{11}
	(U(k),U(k)) \sim k^{-\frac{3}{2}}.
\end{eqnarray}
For the spectrum in Eq.~(\ref{11}) of the quadratic velocity spread, is observed:
\begin{eqnarray}\label{0}
	\u^2(\delta\r) \sim \vert\vert\delta\r\vert\vert^{-\frac{1}{2}}.
\end{eqnarray}

The second term in Eq.~(\ref{helicity4}) can be explained as follows. According to Eq.~(\ref{beta}),
the coefficient $\beta$ is included in the expression for the turbulent magnetic diffusion in Eq.~(\ref{helicity4}), which determines the dissipation rate of a system of magnetic flux tubes.
One can imagine a system of twisted magnetic flux tubes, and then the rate of untwisting of each magnetic flux tube.
Due to turbulent diffusion, which plays the role of a dissipative process for a large scale field, the twist rate is proportional to the coefficient.

The twist coefficient of a magnetic flux tube at each point of its central line is determined by the magnitude of the vector component $\rot \B$ along the line. This coefficient has a scale of the order of the square root of the quadratic spacing of the magnetic vector. Conversely, the mean square small scale spacing of the velocity vector determines the stretch coefficient of each area orthogonal to the magnetic line, which also determines the rate of change of the component $\B$ of the large scale component of the magnetic field, according to the condition of conservation of the vector flux (see Eqs.~(\ref{alpha}) and~(\ref{fluctuation})).
The density of the magnetic helicity invariant, stirred by the fast velocity field, undergoes a variation given by the second term in Eq.~(\ref{helicity4}), where the scaling exponent $\beta$ is determined by the quadratic spread of the energy of small-scale velocity field.

The spacing of each component
of the small scale magnetic field $\b$, for energy reasons, is distributed like $\v$, which determines the distribution of the small scale magnetic field in the Kolmogorov spectrum.
In a MHD system, the large scale magnetic and hydrodynamic fields are in mutual balance, The velocity field transports the magnetic field and the magnetic field influences the velocity field via Lorentz forces. For the magnetic energy spectrum of the large scale field
for the Kolmogorov case in Eq.~(\ref{2/3}), we write the following relation:
\begin{eqnarray}\label{-1}
	U_{\B}(k)=(\B_0(k),\B_0(k)) \sim k^{-\frac{5}{3}},
\end{eqnarray}
which we will call the slope of the Kolmogorov spectrum.

In case in Eq.~(\ref{11}), for the turbulent viscosity spectrum we obtain that $\beta(k) \sim k^{-\frac{1}{2}}$ and
for the current helicity -- $(\B_0,\rot(\B_0)) \sim k^{-\frac{1}{2}}$.

Hence, on the left side, the spectral index for the magnetic helicity flux density is
$(\A_0(k),\B_0(k))^{\rm{flow}} \sim k^{-1}$, whereas the spectral index for the magnetic energy $1D$-flux is $(\B_0(k),\B_0(k))^{\rm{flow}} \sim k^{0}$.
For calculations, we will calculate (conditionally) the magnetic $3D$-flux
\begin{eqnarray}\label{pokaz}
	\frac{\partial \B_0}{\partial t} =	\B_0^{\rm{flow}} \sim k^{-1}.
\end{eqnarray}
The magnetic energy flux index for a magnetic flux index is calculated using the formula $-1 \cdot 2+2=0$.
Similarly, for the Kolmogorov spectrum in Eq.~(\ref{-1}), Eq.~(\ref{pokaz}) is replaced by
\begin{eqnarray}\label{pokaz2}
	\frac{\partial \B_0}{\partial t} =	\B_0^{\rm{flow}} \sim k^{-\frac{7}{6}}.
\end{eqnarray}
We are talking about the spacial scale of the magnetic flux vectors in Eqs.~(\ref{pokaz2}) and~(\ref{pokaz}).

In the calculation, we used the following assumptions.
The magnetic flux is defined by the vector $\B_0^{\rm{flow}} $, and its spatial scale
is reconstructed from the scale of the magnetic energy $(\B_0^{\rm{flow}},\B^{\rm{flow}})$, which coincides with the scale of the magnetic energy flux $(\B_0,\B_0)^{\rm{flow}}$.
The value of the dot product $(\B_0,\B_0^{\rm{flow}})$ is calculated from the value of the flow
$\frac {d\B_0}{dt}$. However the calculated value does not contribute to the value of
$(\B_0,\B_0)^{\rm{flow}}$.

The following method is proposed for estimating the scaling dependence of $(\B_0,\B_0)^{\rm{flow}}$. With a linear variation of the vector $\B_0$ by a factor of $\lambda$, $\lambda \mapsto 1$, due to the stretching of the small scale velocity field, the value of $\frac {d(\B_0,\B_0)}{dt}$ changes by a factor of $\lambda$. At the same time, the linear value of the magnetic vector variation is not caused by the electromotive force in Eq.~(\ref{alpha}) of the mean field and does not lead to a linear contribution to the expression $(\B_0,\B_0)^{\rm{flow}}$.
The area of the transverse plane changes inversely proportionally with the stretching of the magnetic vector.

Let us consider the expression $(\B_0,\B_0)^{\rm{flow}}$ as a quadratic variation of the magnetic energy 
$(\B_0,\B_0)$ of a large scale field. The spatial scales of the quadratic variations caused by the appearance of a component perpendicular to the direction of the magnetic vector and by the longitudinal stretching of the magnetic field vector by the hydrodynamic velocity field
coincide. The spatial index of the magnetic energy flux is calculated from Eq.~(\ref{helicity4}).
Then, the spatial scale of the expression
$(\B_0^{\rm{flow}}(k),\B^{\rm{flow}}(k))$ coincides with the spatial scale of the flux of the square of the length of the magnetic vector $(\B_0,\B_0)^{\rm{flow}}$. This scale corresponds to Eqs.~(\ref{pokaz}) and~(\ref{pokaz2}) for the corresponding MHD spectrum.

For an ideal MHD system, specifying random solutions with spectral densities in the form presented above seems natural. The predicted hydrodynamic indexes determine the slope of the magnetic energy spectrum since free exchange of hydrodynamic and magnetic energy occurs due to the Lorentz force and the magnetic field transfer by the hydrodynamic velocity field.
General considerations for calculating MHD spectral indexes are available in Ref.~ \cite{Z-R-S}.

The density of the Hopf invariant of magnetic lines has the index $a=-1$.
For  the invariant $M_{5}$ the index $a=-\frac{7}{6}$.
We will recall in Sec.~\ref{sec:PROPMAGINV} the calculation for $M_5$ from Ref.~\cite{A4}. This invariant is related with mean field equation (\ref{meaN}). 
A combinatorial formula for $M_5$ is more complicated with respect to the formula (\ref{m}) of $M_3$ and a numerical experiment using two-variables Conway polynomial is in progress.

The existence of a spectral index of the magnetic field that determines the magnetic lines implies the condition that the $M_5$---density is preserved when changing the scale. Thus, the geometric structure of the magnetic lines for such a field, estimated by the invariant value, is independent of the measurement scale. The density index of the invariant corresponds to the spectral index of the large scale magnetic field in Eqs.~(\ref{pokaz}) and~(\ref{pokaz2}).

\section{Properties of magnetic line invariants\label{sec:PROPMAGINV}}

In the dynamo theory, it is generally accepted that the properties of the large scale dynamo are determined by the behavior of the system at a small scale, which is represented by a hierarchy~\cite{Mo,Ra,Z-R-S}. The relationship with the large scale field is described by the mean field equation, discussed above. The small scale field is mixed at each scale by the fast hydrodynamic velocity field of the conducting fluid. The large scale field is the result of averaging the small scale field.

The small scale field is close to an ideal magnetic field, which is transformed like a  magnetic field frozen into a fluid conducting medium. With this approach, the structure of the small scale field can be characterized by the density of magnetic line invariants. These invariants are functions, frozen into the medium in the domain, which depends a priori of a scale. Because a fast small scale velocity field mixes the density of the invariant homohenously, the index of the density has to be independent of a wave-number of magnetic mode. 
With this approach, an exact solution of the MHD equations is out of the question.

The concept of the MHD spectrum, which characterizes the distribution of hydrodynamic and magnetic energy by scale, is defined. In steady state turbulence, a common index for the scale dependence of magnetic and hydrodynamic energy should be specified. In this case, the scale is typically reduced to zero, thereby increasing the wave number to infinity. Of course, the improper integral of the first kind must converge with the MHD index.


Let us assume that, due to mixing, the mean magnetic field invariant density for a steady-state MHD turbulence regime is a scale-independent quantity. It is generally assumed that the rate of magnetic field mixing is so high that the second derivative of the dissipation rate of the magnetic field invariant density becomes very large. In this case, the invariant density approximates the magnetic flux determined by the $\beta$-term in the mean field equation, which is responsible for the magnetic diffusion.
An important role is thus played by the question: ``Do there exist invariants of magnetic lines whose dimension determines a uniform scale distribution of such an invariant with a given value of the magnetic flux index of the MHD spectrum?'' This question is a reformulation of Arnold's problem on the existence of a higher ergodic asymptotic invariant of links \cite{Arn}[problems~[1990-16] and~[1984-12]].

The dimension of an invariant is defined as two real numbers $n_1$, $n_2$, which are written as $G^{n_1}cm^{n_2}$. The index $n_1$ determines how many times the value of the invariant density changes if the magnetic field increases $\mu$-fold? When calculating, it turns out that the index $n_1$ is a natural number; with an increase in the magnetic field, the density of the magnetic line invariant increases. The index $n_2$ determines how many times the value of the invariant density changes if the scale increases $\lambda$-fold? Since we are talking about invariants used for MHD turbulence, the index $n_2$ must be negative. The ratio of the indexes $\frac{n_2}{n_1}$ is called the fractal index of the invariant. The fractal index determines the index of the MHD spectrum for which the invariant is applied to large scale magnetic lines. With this value of the MHD spectrum index, the density of the magnetic line invariant is independent of scale.

The  linking invariant of magnetic lines satisfies the above conditions. The pairwise magnetic line linkage coefficient in our formulation is characterized by a function on the Cartesian square $\Omega \times \Omega$ of the domain with the magnetic field, which is the density function. Such an invariant exists and is called the asymptotic ergodic Hopf invariant, denoted by $\chi$. This invariant was discovered by Arnold in 1974. The word "asymptotic" means that the invariant applies to magnetic fields of a general type, whose magnetic lines are not necessarily closed, but can densely fill a portion of the volume of the domain $\Omega$ or be densely distributed everywhere on the surface.

The word "ergodic" means that the invariant has a density function (since its value is determined by the ergodic theorem applied to magnetic flux) that remains constant during the hydrodynamic transformation of the magnetic field. The fractal dimension of the asymptotic ergodic invariant is $G^2cm^{-2}$, which means that for the large scale component of the Iroshnikov-Kraichnan MHD spectrum with the energy exponent $-3/2$, the density function is scale-independent. It follows that 
the magnetic lines of such a magnetic flux
consist of bundles of pairs of magnetic lines that twist with each other with a constant coupling coefficient. For open magnetic lines, the coupling coefficient is normalized to the magnetic length, which is independent of scale.

The question arises if there are magnetic flux line invariants whose fractal indices correspond to those of MHD turbulence, for example, the Kolmogorov MHD index  $-5/3$. We answer affirmatively. Thus, we assert that the magnetic flux of the Kolmogorov spectrum consists of bundles of quintuplets of magnetic flux lines that twist and intertwine with each other with a constant M-invariant, which is determined not only by the twist coefficient of closely spaced magnetic flux lines, but also by the knotting (intertwining) coefficient of closely spaced magnetic flux lines.

The $M_3$ invariant is constructed in the theory of classical links (a branch of topology) based on the Conway polynomial in one variable. This invariant has the dimension $G^{12}cm^{-6}$ and, therefore, has a fractal index $-1/2$. Such an invariant in magnetic turbulence can be used only to study the large scale magnetic flux field of the MHD spectrum with the index $-1$. If, for simplicity, we assume that the magnetic field consists only of closed lines, then the $M_3$ invariant is expressed combinatorial as the sum of values for all possible different triples of magnetic lines. This is a finite-order invariant in the sense of Vassiliev.

The value of the $M_3$-integral density is determined by the integral convolution over a finite-multiple configuration of points on all possible triples of magnetic lines. This combinatorial formula is given in Sec.~\ref{SEcM}; a numerical experiment confirming the combinatorial formula is given in Sec.~\ref{SecM2}. The fractal density is calculated in the limit by applying Birkoff's ergodic theorem to the magnetic flux that generates auxiliary magnetic fluxes on the configuration space.

For the simpler case of quadratic magnetic helicity \cite{A3}, it has been proven that the invariant is ergodic for magnetic fields of smoothness class $C^{2}$, i.e., for fields whose components have continuous second partial derivatives. The smoothness class can probably be lowered (for smoothness class $C^{1}$, the statement is false); this problem has not been fully investigated. It is interesting that in the context of Theorem in Ref.~\cite{K} the smoothness class of the quadratic magnetic helicity invariant cannot be lowered to class $C^1$, since, obviously, the distribution of the second moment of the magnetic line linking coefficient (the asymptotic ergodic Hopf invariant) depends on the magnetic field configuration.

The invariant $M_5$, suitable for describing the magnetic flux of the Kolmogorov MHD spectrum, is defined by the corresponding integral over the configuration space constructed from quintets of magnetic lines \cite{A4}. When defining $M_5$, one must assume that the quintet of magnetic lines is equipped with a fixed cyclic order. One can visually consider the continuum of magnetic lines to be colored by points on a standard circle $S^1$, with the colors uniformly fractolized by scale. This kind of hypothesis is a mathematical abstraction that allows one to correctly define the asymptotic ergodic integral.

From a physical perspective, the existence of $S^1$-colors of magnetic lines has not been confirmed at this stage. However, it is possible to define a symmetrization and, for each configuration of five magnetic lines with a prescribed cyclic order, determine the value of all possible re-colorings of the magnetic lines in a given configuration. We will call this symmetrization "loss of internal  asymmetry".

As a result, the invariant $M_5$ is simplified; it turns out that its value can be completely calculated from the values of $M_3$ from all possible triplets of magnetic lines in the selected five and from all possible pairwise coupling coefficients (the choice of both triplets and pairs in proper sub-configurations can be made unordered). This occurs because, when asymmetry is forgotten, the combinatorial invariant $M_5$, constructed on the basis of the Conway polynomial of two variables, is simplified into an invariant constructed on the basis of the Conway polynomial of one variable, and such an invariant does not remember the values of the invariant calculated over all possible ordered quadruples of magnetic lines included in the quintuple of cyclically ordered magnetic lines. If we neglect  contributions of quadruples of knotted magnetic lines, then the density of the $M_5$-invariant loses its geometric meaning, since it is expressed as the product of densities of polynomial momenta of linking numbers and the density of $M_3$-invariant, lifted to the configuration space of quintuples of magnetic lines.

\subsection*{Asymptotic ergodic Hopf invariant}

We consider smooth magnetic fields without vanishing points in a compact domain, undirected along the tangent to the domain boundary. The simplest domain of this type is a solid torus with a boundary torus; such a domain is called a magnetic tube. Magnetic flux is a one-parameter family of incompressible diffeomorphisms of the domain onto itself, such that points in the domain move along magnetic lines with velocities prescribed by the magnetic field. 
We define magnetic lines in $\Omega$ as integral trajectories of the magnetic field $\B$.

At the boundary surface of the domain, the trajectories are the windings of the boundary tori. In the case of a magnetic flux tube, the magnetic lines inside it form parallel windings along a central closed line, which is oriented in the direction of the magnetic lines. We do not assume that the magnetic lines are closed either inside or at the boundary of the magnetic flux tube. In the case of integrable magnetic flux dynamics, the $\Omega$ domain is layered into concentric tori around a closed invariant magnetic line.

Under a small and sufficiently smooth perturbation of such a system, the family of tori is not destroyed, but only slightly deformed into a countable family of invariant tori, which serve as the boundary surfaces of concentric cylindrical regions, within each of which any magnetic line is everywhere dense and its position is completely characterized by a finite set of angular frequencies. However, in the more general case, far from integrable, the dynamics of magnetic flux are little studied. In this case, the magnetic field is completely characterized by a Poincar\'{e} diffeomorphism, which maps the central cross-section of the magnetic tube onto itself.

An invariant of the magnetic field $\B$ under the action of the group $DiffS(\Omega)$ of volume-preserving coordinate changes is a function on $\Omega$, calculated from $\B$ and unchanged by the group action. The main example of a magnetic field invariant is the asymptotic ergodic Hopf invariant, which is denoted by $\chi$. In the case where the magnetic field has only closed magnetic lines, the invariant $\chi$ is defined as follows.

Let us choose an arbitrary point $(x_1, x_2)$ in $\Omega \times \Omega$ and consider the magnetic lines $L( x_1), L( x_2)$ (nodes) through the chosen points. Let $T_i$ denote the value of the magnetic parameter that ensures a single traverse along the magnetic line $L_i$ (it is clear that $T_i$ does not depend on the point on the magnetic line). Let us determine the coupling coefficient $k(L_1, L_2)$ of the magnetic lines using the equation:
\begin{eqnarray}\label{1}
	k(L_1, L_2) = T_1^{-1}T_2^{-1}k(L_1, L_2).	
\end{eqnarray}
Let us define the asymptotic Hopf invariant using the equation:
\begin{eqnarray}\label{2}
	\chi(x_1, x_2) = k(L_1(x_1), L_2(x_2)).	
\end{eqnarray}
In the case of a general magnetic field, equation (\ref{2}) should be understood in the asymptotic limit as $T_1,T_2 \to +\infty$ based on the individual ergodic theorem \cite{H}. The coupling coefficient of magnetic line segments of magnetic lengths $T_1$, $T_2$ from a prescribed pair of points is calculated by the Gauss integral (see \cite[Chapter III, 4.3]{A-Kh}). In this case, the function given by (\ref{2}),  defined almost everywhere and Lebesgue integrable, represents an element of the space of absolutely integrable functions in the domain $\Omega \times \Omega$.

The density function has dimensions $G^2cm^{-2}$. A convolution of this function with respect to one (the first) argument is used, which is called the magnetic helicity density. The magnetic helicity density is also invariant under transformations of the domain with a magnetic field.

Since the density function is an invariant of the magnetic field under $DiffS(\Omega)$, we can consider not only the average value of this function, which is called the magnetic helicity integral, but also higher integral moments. A pair of different second moments is called the quadratic magnetic helicity integral \cite{A3}. It is important to note that within one ergodicity region, where the magnetic line is everywhere dense, the magnetic field invariant reduces to a number equal to the Hopf invariant of the magnetic line \cite{B-M}. As a rule, the number of ergodic regions is countable, so the quadratic magnetic helicity invariant over the entire magnetic field region does not reduce to the magnetic helicity invariant, although it does reduce to the magnetic helicity invariant in each subregion where the magnetic field is dense everywhere. The quadratic helicity invariant is applied in Refs.~\cite{S,D,A-D2}.

\subsection*{Integral $M_3$} 
The integral invariant $M_3$, constructed in Ref.~\cite{A2}, is defined only for a magnetic field with closed magnetic flux lines. The terms of this integral are described in Theorem 6, p. 8 \cite{A2}.
For a magnetic field of this type, $M_3$ is defined as the integral value of all possible triples of magnetic flux lines. If, for each triple of magnetic flux lines, we extend it to a triple of infinitely thin magnetic tubes with normalized magnetic flux through a transverse cross-section, then the integral is defined by an analytical expression. This expression is used below to calculate the dimension of the $M_3$-integral. The condition of closed magnetic lines can be removed by Arnold's argument, applying the Ergodic Theorem for fluxes on the configuration space. Inside the only ergodic domain Baader-March\'{e} Theorem \cite{B-M} is not satisfied as can be deduced from the combinatorial equation (\ref{m}). The orders of the right-hand side of the equation, including the polynomial term, equals to $7$, but the
asymptotic degree in the formula
(\ref{ass}) equals to $12$: $7\cdot2 \neq 12$.

The integral $M_3$ has terms (14)-(16) p. 4; (24) p. 8; Lemma 7, p. 9, for which the ergodicity property without the assumption of closed magnetic lines is not obvious and requires proof. For example, such a term is presented in \cite{A2}, Fig. 2, p. 11, Example 12. 
The ergodicity of this and other terms of a similar type in $M_3$ is clear without the assumption that the magnetic lines are closed. The integral contains terms of a different type. We examine one such term in Appendix I, Fig. 2. Earlier in publication \cite{A8}, the first author encountered difficulties in attempting to prove the ergodicity.
It is easy to see that when expanding all quantities in a Fourier series, the problem of small denominators arises, which can be overcome within the framework of KAM theory. In our special case, we can carry out a construction that does not use KAM theory, without going beyond the short-interval technique \cite{A-Kh} (Chapter III, 4.3, p. 163).

\subsubsection*{Dimension of $M_3$}
The dimension of the principal term of the invariant $M_3$ is $G^{12}cm^{-6}$ \cite{A2}.
The dimensions of the remaining terms are calculated similarly.
Along magnetic field lines, a magnetic flux is defined for which the ergodic theorem can be applied and the integral kernel values can be averaged. Averaging along a magnetic field line does not change the dimension of the quantity. This averaging results in a density distributed in the Cartesian cube $\Omega^3$ of the domain with the magnetic field. For non-contracting diffeomorphisms of the $\Omega$ domain, the density function transforms invariantly.

The fractal index of the $M_3$ invariant is $\frac{6}{12}=\frac{1}{2}$.
It is important that the $M_3$ invariant is defined without the assumption that the Hopf invariant density is of constant sign. In the $\Omega^2$ phase space, there can exist both ergodic zones in which pairs of magnetic lines are linked with a positive coefficient and ergodic zones in which the linking coefficient is negative.

\subsection*{Arnold's inequality and its generalization}
Arnold's inequality is a lower bound for the magnetic energy in a compact region with a fluid conducting medium in terms of magnetic helicity, \cite{A-Kh}.
The integral $M_3$ does not, generally speaking, define any explicit lower bound for the magnetic energy, since it has an order  $12$ with respect to  magnetic fields. For the simpler case of quadratic magnetic helicity (the second moment of the coupling coefficient of magnetic lines), as shown in \cite{A-D2}, such an estimate is impossible; it is only possible to estimate from below the $L^4$-norm of the magnetic vector.

Of course, the invariants of the leading moments of the pairwise magnetic line coupling coefficients \cite{A-V} also estimate the distribution of the leading moments of magnetic energy, which provides implicit information about its distribution.

\section{Equation of the $M_3$-invariant}\label{SEcM}

For $m=3$, consider an oriented ordered link 
$\L = L_1 \cup L_2 \cup L_3$. For such a link, the Melikhov invariant $\gamma(\L)$ is defined as in (cf. \cite{A2} Theorem 17):

$$
\gamma(\L) = c_1(\L) 
$$
$$-((1,2)(2,3) + (2,3)(3,1) +
(3,1)(1,2))(c_1(L_1) + c_1(L_2) + c_1(L_3))$$
$$ -((3,1)+(2,3))(c_1(L_1 \cup L_2) - (1,2)(c_1(L_1) + c_1(L_2))) $$
$$ -((1,2)+(3,1))(c_1(L_2 \cup L_3) - (2,3)(c_1(L_2) + c_1(L_3))) $$
$$ -((2,3)+(1,2))(c_1(L_3 \cup L_1) - (3,1)(c_1(L_3) + c_1(L_1))), $$
where $(i,j)=c_0(L_i \cup L_j)$ denote the link coefficients of pairs of components $L_i$, $L_j$, $i,j = 1,2,3$, $i \ne j$,
$c_1$ is the coefficient in the Conway polynomial 
$\nabla_{\L}(z)=z^{m-1}(c_0 + c_1z^2 + \dots )$
for the corresponding proper $m$-component sub-link, $m=1,2,3$.
The Conway polynomial is used as in 
\cite{P-S}.

Let us define the invariant
$\tilde M(\L)$ by the equation:
\begin{eqnarray}\label{MInv}
	\begin{array}{c}
		\tilde M(\L) = -(1,2)(2,3)(3,1)\gamma(\L) + \\
		((1,2)^2(2,3)^2\beta(L_3 \cup L_1) + (2,3)^2(3,1)^2\beta(L_1 \cup
		L_2) + (3,1)^2(1,2)^2\beta(L_2 \cup L_3)), 
	\end{array}
\end{eqnarray}
where
\begin{eqnarray}\label{SatoLevine}
	\beta(L_i \cup L_j) = c_1(\L) - c_0(\L)(c_1(L_1) + c_1(L_2)) 
\end{eqnarray}
is the Sato-Levin invariant for an oriented 2-component link
$\L = L_i \cup L_j, \quad i \ne j$.

The following theorem is proved in the proceeding
\cite{A-S}. 

\subsubsection*{Theorem 1}

1. There exists an odd polynomial $P((2,3),(3,1),(1,2))$ in pairwise linking coefficients. The equation for the polynomial under the assumption $(1,3)=(2,3)$ is written below, 
such that
\begin{eqnarray}\label{m}
	M(\L) = \tilde{M}(\L) + P((2,3),(3,1),(1,2))
\end{eqnarray}
is asymptotic. This means that for an arbitrary set of pairwise coupling coefficients
$(3,1),(1,2),(2,3)$ (under the assumption above) and for an arbitrary vector with nonzero coordinates $(\lambda_1,\lambda_2,\lambda_3)$ (under the assumption $\lambda_2 = \lambda_3$), which is called the coil vector, the invariant $M$
satisfies the relation:
\begin{eqnarray}\label{ass}
	M(\lambda_1 L_1 \cup \lambda_2 L_2 \cup \lambda_3 L_3) = \lambda_1^4 \lambda_2^4 \lambda_3^4 M(L_1 \cup L_2 \cup L_3). 
\end{eqnarray}
(the operation  $L_1 \cup L_2 \cup L_3 \mapsto \lambda_1 L_1 \cup \lambda_2 L_2 \cup \lambda_3 L_3$,
$\lambda_i \in \Z_+$, 
$i=1,2,3$ is called the cable operation). 
The polynomial $P$ is the following: $P=P_1+R$, where 
\begin{eqnarray}\label{P1}
	\begin{array}{c}
		P_1((2,3),(3,1),(1,2))=               \\ 
		+\frac{1}{6}(1,2)^2(2,3)^2(3,1)^2[(1,2)+(2,3)+(3,1)]  \\       
		+\frac{1}{12}(1,2)(2,3)(3,1)[(1,2)(2,3)+(2,3)(3,1)+(3,1)(1,2)]. 
	\end{array}
\end{eqnarray}

The general case of three different pairwise coupling coefficients turned out to be more complex, and a numerical experiment in this case was not conducted.
We present the equation for the $R$-term of the polynomial in the case $(1,3)=(2,3)$:
\begin{eqnarray}\label{polR}
	R=1/12(1,3)^6(1,2) + 1/6(1,3)^4(1,2)^3   
\end{eqnarray}
Equation in Eq.~(\ref{polR}) 
should be considered for an arbitrary linkage (under the condition $0 \le (1,2)(2,3)(3,1)=k$, we obtain a right linkage with $(1,2)>0$, and under the condition $0 \ge (1,2)(2,3)(3,1)=k$, we obtain a left linkage with $(1,2)<0$).

2. The combinatorial invariant on the left-hand side of equation in Eq.~(\ref{m})
coincides with the integral invariant constructed in \cite{A2}.

3. The invariant $M$ in formula in Eq.~(\ref{m})
is skew-mirror or mirror-asymmetric. This means that for a symmetric link (mirror symmetry of space is defined as a symmetry transformation relative to the chosen projection plane, which also can change the cyclic order of the link components $\L$)
the invariant's value is reversed:
$M \mapsto -M$.

\[  \]

\section{Equation for $M_5$-invariant}

Let us recall the combinatorial equation (33) \cite{A4}, Theorem 16, with corrections (there is a misprint in it).
Let $\L=(L_1,L_2,L_3,L_4,L_5)$ be an arbitrary $5$-component oriented ordered link. Assume that all pairwise linking numbers between components of $\L$ are
nontrivial: $k(i,j) \neq 0$, $1 \le i < j \le 5$. Then define
{\small  
	\begin{eqnarray}\label{M5comb}
		\begin{array}{c}
			\\
			5 M_5 = \\
			\sum_{ciclic} \Delta_{(1,2);(3,4)} (1,2)(3,4)(1,5)(2,5)(3,5)(4,5) + \Delta_{(1,3),(2,4)}(1,3)(2,4)(1,5)(2,5)(3,5)(4,5) + \\
			\overline{\gamma}_{1,2,3,4}  + \overline{\beta}_{1,2,3,4} + 	S((1,2)(1,3)(1,4)(1,5)(2,2)(2,4)(2,5)(3,4)(3,5)(4,5)). \\
		\end{array}	
\end{eqnarray}}		
where
{\small
	\begin{eqnarray}\label{M5combG}
		\begin{array}{c}	
			\\	
			\overline{\gamma}_{1,2,3,4}=
			\\ \gamma_{1,2,3}(1,4)(1,5)(2,4)(2,5)(3,4)(3,5)(4,5) + 	\gamma_{1,2,4}(1,3)(1,5)(2,3)(2,5)(4,3)(4,5)(3,5) + \\
			\gamma_{1,3,4}(1,2)(1,5)(3,2)(3,5)(4,2)(4,5)(2,5) + \gamma_{2,3,4}(2,1)(2,5)(3,1)(3,5)(4,1)(4,5)(1,5);
		\end{array}	
\end{eqnarray}}
{\tiny
	\begin{eqnarray}\label{M5combB}
		\begin{array}{c}			
			\overline{\beta}_{1,2,3,4}= \\
			(\beta_{1,2}(1,3)(2,3)(1,2)^{-1} + \beta_{2,3}(1,2)(3,1)(2,3)^{-1} + \beta_{3,1}(3,2)(1,2)(1,3)^{-1})(1,4)(1,5)(2,4)(2,5)(3,4)(3,5)(4,5) + \\
			(\beta_{1,2}(1,4)(2,4)(1,2)^{-1} + \beta_{2,4}(2,1)(4,1)(2,4)^{-1} + \beta_{4,2}(4,1)(2,1)(4,2)^{-1})(1,3)(1,5)(2,3)(2,5)(4,3)(4,5)(3,5) + \\ 
			(\beta_{1,3}(1,4)(3,4)(1,3)^{-1} + \beta_{3,4}(3,1)(4,1)(3,4)^{-1} + \beta_{4,1}(4,3)(1,3)(4,1)^{-1})(1,2)(1,5)(3,2)(3,5)(4,2)(4,5)(2,5) + \\
			(\beta_{2,3}(2,4)(3,4)(2,3)^{-1} + \beta_{3,4}(3,2)(4,2)(3,4)^{-1} + \beta_{4,2}(4,3)(2,3)(4,2)^{-1})(2,1)(2,5)(3,1)(3,5)(4,1)(4,5)(1,5). \\
		\end{array}	
\end{eqnarray}}							

In the equation (\ref{M5comb})  the sum  is taken over 5 terms:
each cyclic permutation of the indexes 
$(1,2,3,4,5)$; $\Delta_{(i,j);(k,l)} = C^{2,2}(L_i \cup L_j,L_k \cup L_l)$ is the coefficient of the Conway 2-variable polynomial, $S$ is an unknown polynomial of pairwise linking numbers.

By the symmetrization of the equation (\ref{M5comb}) we get the following equation, which for the link $\L$ with no order of components: 
\begin{eqnarray}\label{su}
	\begin{array}{c}
		M^{\rm{sim}}_5 = \\
		\sum (l,m)(l,i)(l,j)(l,k)(m,i)(m,j)(m,k)(i,j)^{-1}(j,k)^{-1}(k,i)^{-1}M_3(i,j,k), 
	\end{array}
\end{eqnarray}
the sum is tacken over $C_5^3=10$ $(i,j,k)$-terms.

The Milnor's identity 
\begin{eqnarray}\label{Miln}
	\mu_{(1,2);(2,3)}+\mu_{(1,3);(2,4)}+\mu_{(1,4);(2,3)}=0
\end{eqnarray}
is generalized for the symmetrization $M_5 \mapsto M^{\rm{sim}}_5$ invariant (\cite{A2}).
The main term in (\ref{M5comb}) is expressed Conway 2-variable polynomial.

We describe the equation (\ref{M5comb}) numerically, using a trick with the recursion of the one-variable Conway polynomial to describe two-variable Conway polynomial. The coefficient $C_{2,2}$ is described using two-steps (two one-steps for each uniformly colored pair of components) using the skein-identity for $C_{2,2}$ from $C_{1,1}$.  For the associated 2-component link with two pairs of colored components the result of the skein-relation is the $4$-term alteration. The alteration  is for two points (the alteration is the signed sum of $4$ terms, associated with the discriminant of two self-intersection points of the link), on the intersections of each pair of uniformly colored components. In the case  magnetic lines in a quintuple are unknotted, the coefficients $C_{1,1}$ in two-variable coincide with $C_2$ in one-variable Conway polynomials, because of the relationship: $C_2 = C_{2,0} + C_{1,1} + C_{0,2}$. In the case the components are unknotted we get only the middle term in the right-hand side of the relationship. The numerical experiment in two variables is not investigated. Thus, the polynomial  $S$ in the equation (\ref{M5comb}) for two-variable is unknown. We conjecture that using the presented recursion we not only may express the main term using one-variable Conway polynomial, but also get the polynomial $S$ as the straightforward extension of (\ref{m}) by  the equation (\ref{su}).

This is demonstrated by the numerical experiment presented in the Section \ref{SecM3}.
The internal asymmetry corresponds to the spectral index, for which the spacial distribution of 
the kernel (\ref{M5comb}) into (\ref{su}) is dimensionless (is uniformity distributed over the spacial scale).  
The invariant in (\ref{M5comb}) is ergodic. The calculation
of each prescribed 5-tuple lines gets 3 different cyclic sub-configurations. The sum of (\ref{M5comb}) over the 3 configurations determines the kernel by the expression (\ref{Miln}). We may apply the invariant (\ref{su}) and take the index of the invariant, such that the random values of the invariant in the kernel using the identity (\ref{Miln}) for each $4$-tuple inside the $5$-tuple. This means that the scaling of the
internal asymmetry (with a random cyclic order of quintuples magnetic lines) determines the index of the value (\ref{su}), which coincids with the integral \cite{A2}. The combinatorial equation of the invariant $M_3$ for closed magnetic lines is in Section \ref{SEcM}. The equation (\ref{M5comb}) coinsids with the integral \cite{A4}. The index of the invariant (\ref{su}) by the lost of internal asymmetry coincides with the index of magnetic  flux of the  Kolmogorov spectrum \cite{A4}, \cite{A9}.


\section{Experimental verification of
	Theorem 1 
}\label{SecM2}

Example 1.

\begin{figure}
  \centering
  \includegraphics[scale=0.5]{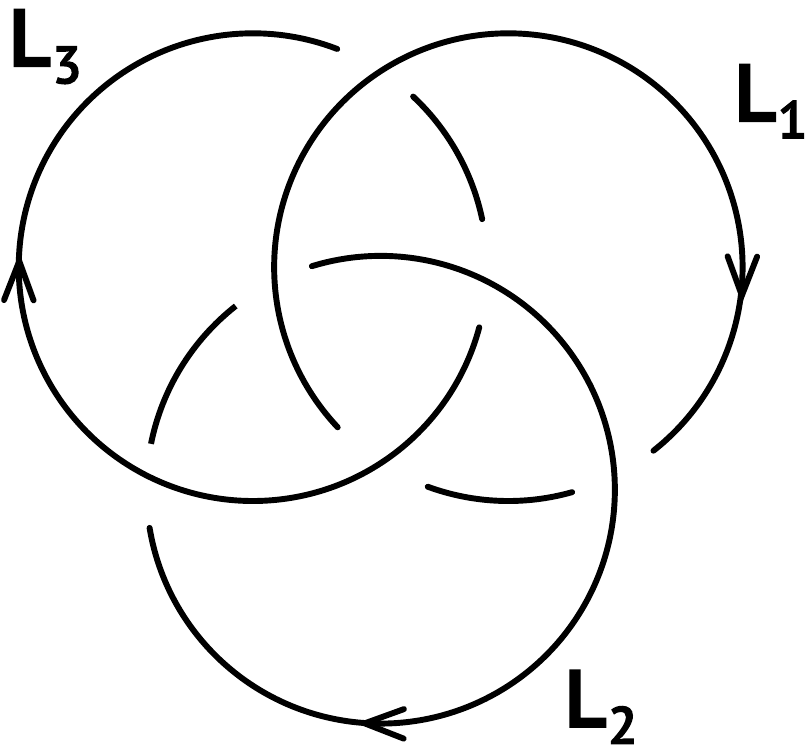}
  \protect
\caption{The Hopf link.}\label{fig:1}
\end{figure}

Consider the Hopf link $\L_{\rm{Hopf}}(+1,+1,+1) = (L_1 \cup L_2 \cup L_3)$, presented in Fig.~\ref{fig:1}.

For this link we have:
$(1,2)=(2,3)=(3,1)=+1$;  $C_1=1$, $C_1(1,2)=C_1(2,3)=C_1(3,1)=0$, $\gamma=1$; $\beta_{2,3}=\beta_{3,1}=\beta_{1,2}=0$; $\tilde{M}=-1$;
$P_1=+\frac{3}{4}$, $R=-\frac{1}{3}$, $M=0$.
\[  \]

Example 2.

\begin{figure}
  \centering
  \includegraphics[scale=0.5]{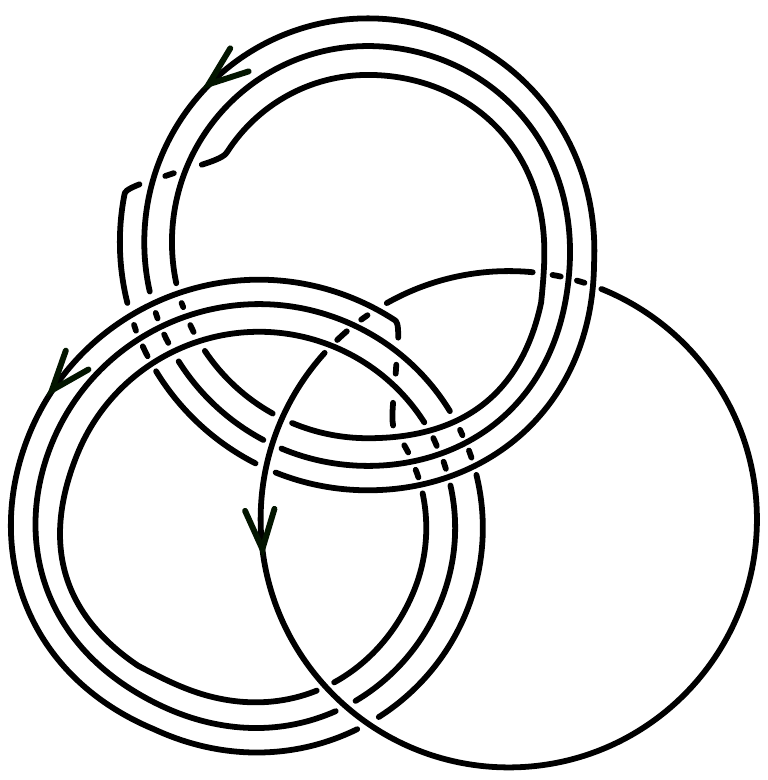}
  \protect
\caption{The (3,3,1)-cable over the Hopf link.}\label{fig:2}
\end{figure}

Consider the link $( 3L_1 \cup 3L_2 \cup L_3)$,  associated with the link $\L_{\rm{Hopf}}(+1,+1,+1)$, see Fig.~\ref{fig:2}.

\[  \]
We have: $(1,2)=(3,1)=+3$, $(2,3)=+9$; $C_1=504$, $\beta_{1,2}=24$, $\beta_{3,1}=0$, $\beta_{2,3}=0$; $\gamma = 360$; $\tilde{M}=-7\cdot2^4\cdot3^5$, $P_1=\frac{277\cdot3^5}{4}$, $R=8748+1640,25$, $M=0$.
\[  \]

Example 3.

Consider the link $(2L_1 \cup L_2 \cup L_3)$,  see Fig.3,  associated with the special link $(1,2)=2$, $(3,1)=3$, $(2,3)=6$ (Fig.5).

\begin{figure}
  \centering
  \includegraphics[scale=0.5]{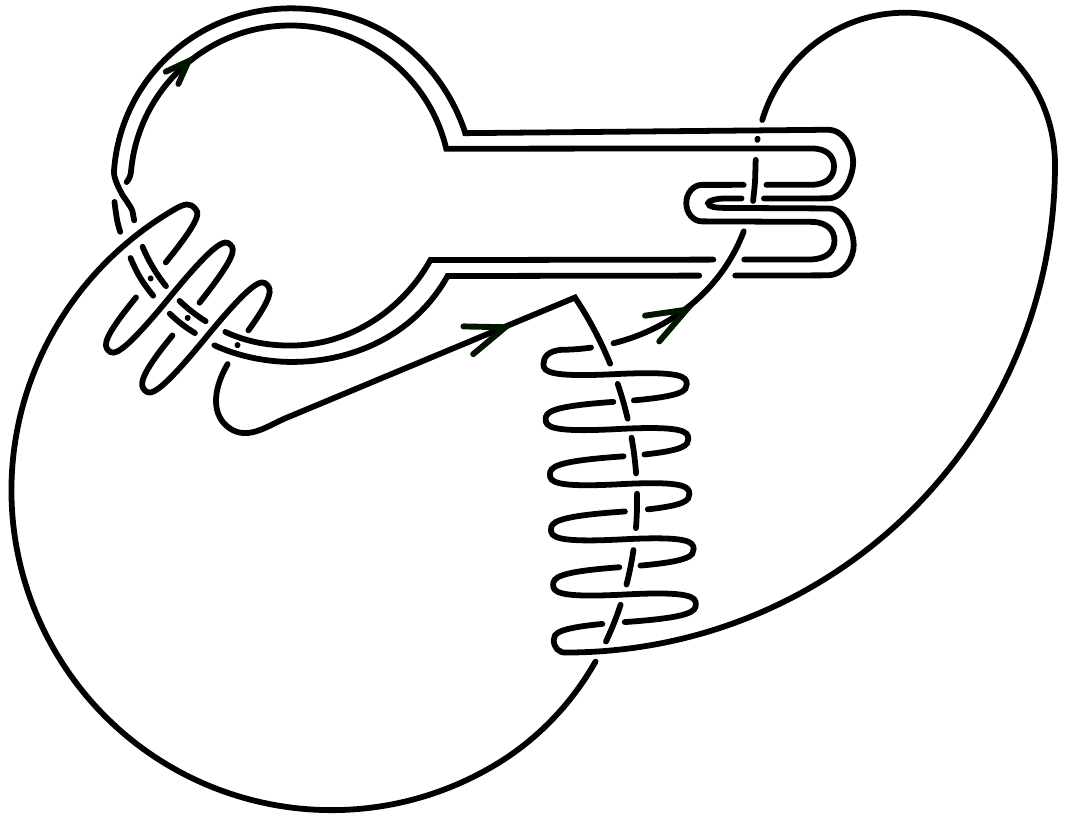}
  \protect
\caption{The (2,1,1)-cable over Fig.~\ref{fig:5}.}\label{fig:3}
\end{figure}

\[  \]
For $(2L_1,L_2,L_3)$ we have 
$(1,2)=4$, $(2,3)=6, (3,1)=+6$; $C_1=1032$, $\beta_{1,2}=1$,$\beta_{3,1}=5$, $\beta_{2,3}=0$; $\gamma = 180$; $\tilde{M}=-21744$, $P_1=56304$, $R=24192-3^4\cdot2^6$; $M=58752-3^4\cdot2^6$.
\[  \]

Example 4. 

\begin{figure}
  \centering
  \includegraphics[scale=0.5]{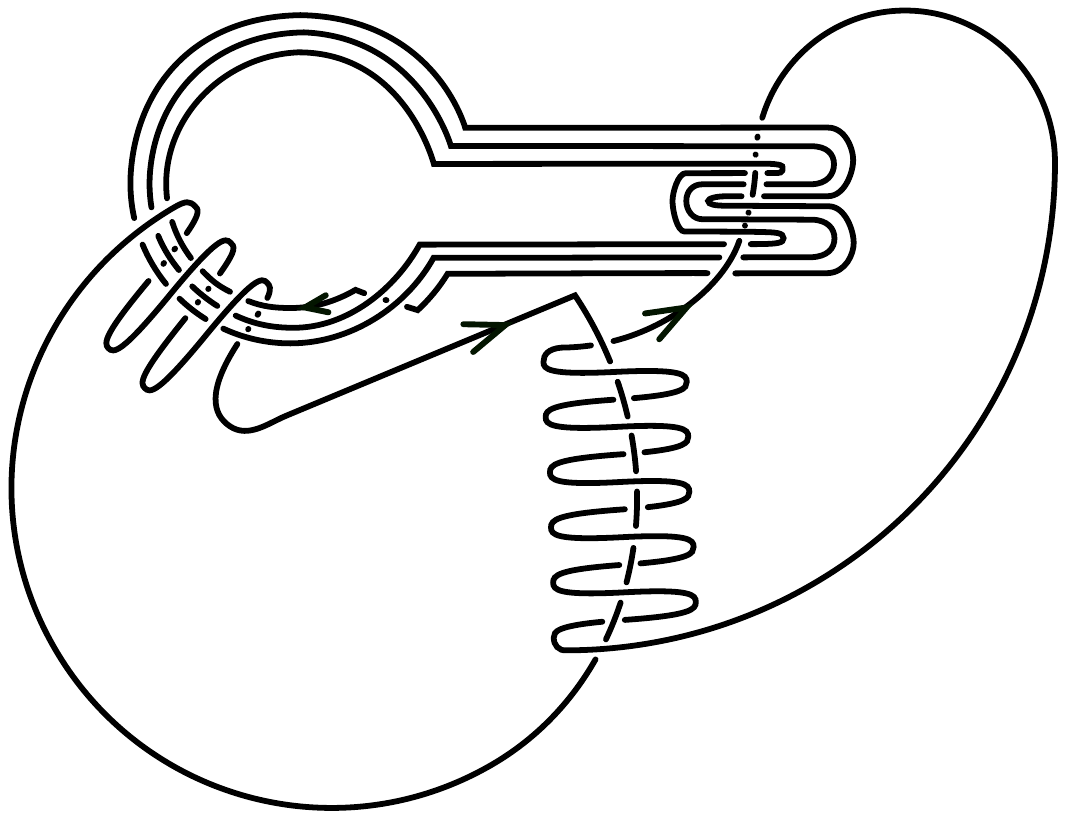}
  \protect
\caption{The (3,1,1)-cable over Fig.~\ref{fig:5}.}\label{fig:4}
\end{figure}

\begin{figure}
  \centering
  \includegraphics[scale=0.5]{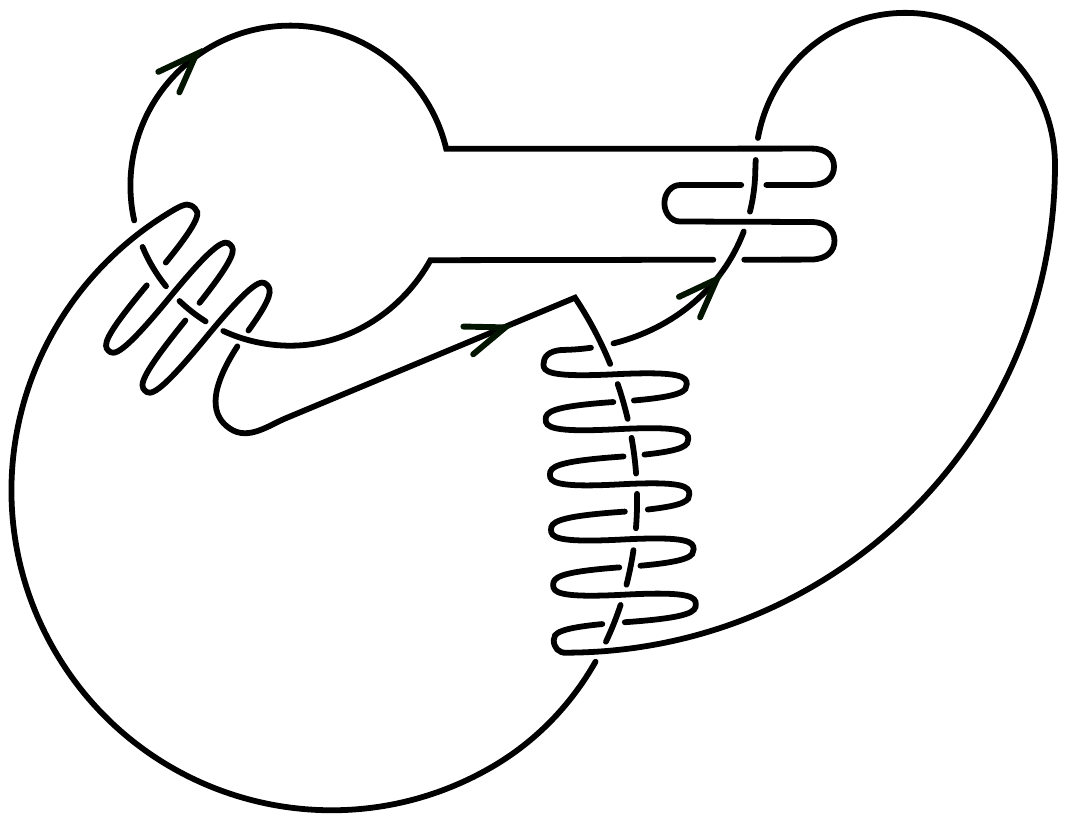}
  \protect
\caption{The base link for Figs.~\ref{fig:3} and~\ref{fig:4}.}\label{fig:5}
\end{figure}

Consider the link $(3L_1 \cup L_2 \cup L_3)$,  see Fig.~\ref{fig:4},  associated with the special link (see Fig.~\ref{fig:5}).

For $(2L_1,L_2,L_3)$ we have 
$(1,2)=6$, $(2,3)=9, (3,1)=+6$; $C_1=1245$, $\beta_{1,2}=5$,$\beta_{3,1}=24$, $\beta_{2,3}=0$; $\gamma = 882$; $\tilde{M}=-240084$, $P_1=371304$, $R=166212-3^8\cdot2^2$; $M=297432-3^8\cdot2^2$.
We get: $3^4M(Fig.4) = 2^4M(Fig.3)$.

The calculations in Examples 1 and 2 disprove the conjecture in \cite{A2}, according to which
the value of the $M_3$-invariant on the Hopf link $\L_{\rm{Hopf}}$ is nonzero. The calculation corrects an error in \cite{A7}.

In papers \cite{A-D} and \cite{D-A2}, it is used that the value of the large scale $\beta$-term for the Chern-Simons wave in the expanding universe model can be described by the Kolmogorov spectrum for physical reasons. This reasoning contained an inaccuracy due to the unknown value of the residual polynomial in the equation in Eq.~(\ref{m}) for the $M_3$-invariant. Since the magnetic flux (the Kolmogorov spectrum) is characterized by the $M$-invariant, such a flux can be visualized in the 3D model under consideration. It is noted that the sectional curvature parameter of the model influences the value of the $M$-invariant of the magnetic flux.

Indeed, as the sectional curvature of the horizontal cross-section decreases, the intertwined bundle of magnetic lines of characteristic width (the bundle width characterizes the contribution of the pairwise line commutators to the M-invariant) increases, and the characteristic bundle length increases. The condition of preserving the M-invariant value for the left bundle leads to a right-handed unwinding of the bundle. Thus, the M-invariant value is related to the Hopf invariant flux and, consequently, to the magnitude of the alpha effect \cite{Ra}. Consequently, the magnitude of the alpha effect depends on the sectional curvature, taking into account Kolmogorov's law. 
The main conclusions of papers \cite{A-D} and \cite{D-A2} remain unchanged.

\section{Experimental verification: the lost of internal asymmetry }\label{SecM3}

\begin{figure}
  \centering
  \includegraphics[scale=0.2]{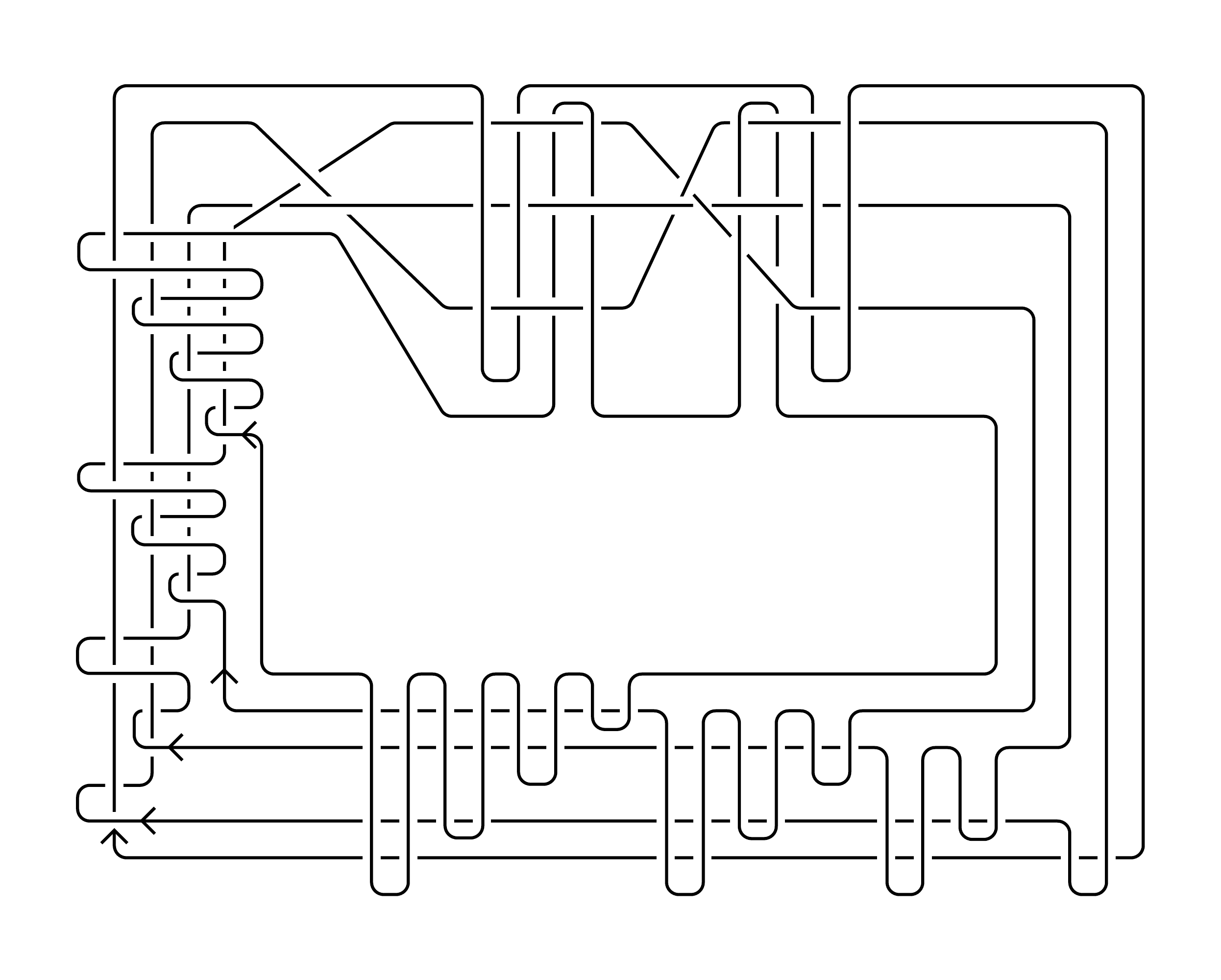}
  \protect
\caption{The braid before a balanced $\Delta$-move.}\label{fig:6}
\end{figure}

\begin{figure}
  \centering
  \includegraphics[scale=0.2]{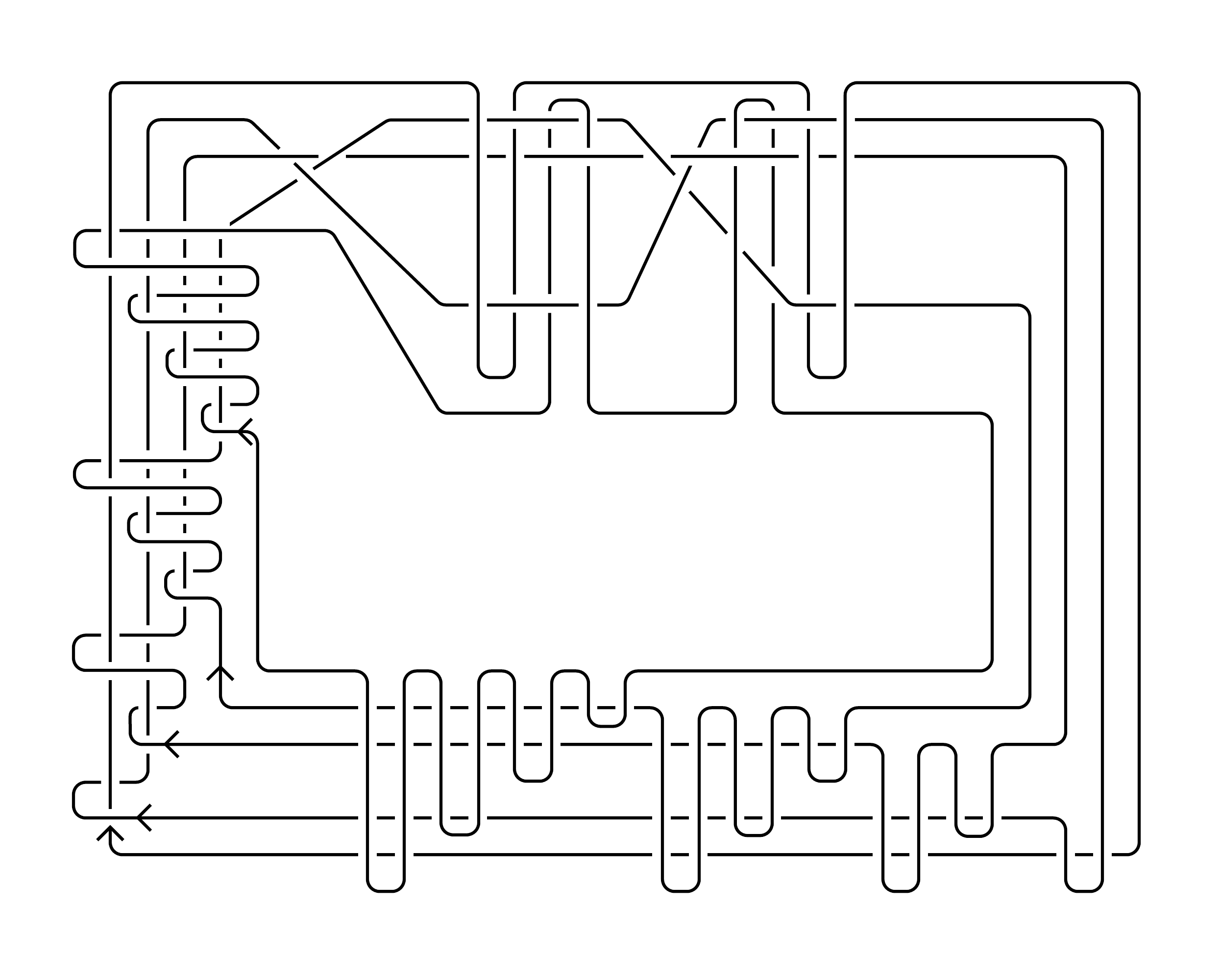}
  \protect
\caption{The braid after a balanced $\Delta$-move.}\label{fig:7}
\end{figure}

The $5$-lines configurations are presented in Figs.~\ref{fig:6} and~\ref{fig:7}. The configurations contain no
knotted lines, all $\beta$ for each pair of lines  equal to zero.

 The magnetic lines in the configurations are ordered by $\{1,2,3,4,5\}$ down-up using the bottom horizontal fragment.
This we get the following 3 cyclic configurations: $\{1,2,3,4,5\}$, $\{2,3,1,4,5\}$, $\{3,1,2,4,5\}$ that keeps the sign of the permutation. With the each permutation the following pair of ordered configurations is associated: 
$$\{1,2,3,4,5\} \mapsto \{1,2,3,4,5\} \cup \{1,3,5,2,4\};$$
$$\{2,3,1,4,5\} \mapsto  \{2,3,1,4,5\} \cup \{2,1,5,3,4\};$$
$$ \{3,1,2,4,5\} \mapsto \{3,1,2,4,5\} \cup \{3,2,5,1,4\}.$$
Inversions of the cyclic order and  odd permutations are not considered.

The each pair, namely the pair $\{1,2,3,4,5\} \cup \{1,3,5,2,4\}$, is associated with
$10$ $2+2$ portion of lines as following: $(1,2),(3,4)$, $(2,3),(4,5)$, $(3,4),(5,1)$, $(4,5),(1,2)$, $(5,1),(2,3)$; $(1,3),(5,2)$, $(3,5),(2,4)$, $(5,2),(4,1)$, $(2,4),(1,3)$.
Each $2+2$ portion determines the value of the main $\Delta$-term in (\ref{M5comb}).

The $\Delta$ term for each prescribed partition is detected by, so-called, balanced $\Delta$-move, where the each of 2 symmetric triangles of magnetic lines (Fig.8) is simultaneously deformed into opposite triangle (Fig.9). The balanced $\Delta$-moves is detected
the Milnors's $\mu_{1,2,3,4}$ invariant, as was proposed in \cite{A9}.

For each configuration only two 4-tuples of lines, 3 lines are organized into the pair of triangles, which are translated by the balanced $\Delta$-move (Fig.~\ref{fig:6} $\mapsto$ Fig.~\ref{fig:7}). The last line of the triangle is variates. The each 4-tuple is organized into 3 $2+2$-tuple and the each $2+2$-tuple is described by the skein-relation of the corresponding balanced-$\Delta$-move using
the alteration of the corresponding prescribed pairs of lines. The alteration is defined by the
intersection of the vertical upper "finger" with the corresponding second horizontal line on the bottom horizontal configuration of lines. The each alteration contains one-variable Conway,  $4$-polynomial, which presented on the two tables. 

\begin{equation}\label{Fig6up}
	\begin{array}{cccc}
		Fig.~\ref{fig:6}  &  A=(1,2);B=(3,4)             &  A=(1,3);B=(2,4)          &  A=(1,4);B=(2,3)   \\
		+origine;up &  \dots - 47z^5 - 16z^3 & \dots -47z^5 -16z^3   &  \dots - 47z^5 - 16 z^3 \\ 
		-A  & \dots - 90z^5  - 24 z^3 & \dots -82z^5  - 24z^3  &  \dots  -95z^5  - 24 z^3 \\
		-B  &  \dots  -75z^5 - 24z^3  &  \dots  -81z^5  - 24z^3 & \dots -86z^5 - 24z^3 \\
		+A,+B & \dots -142z^5 - 36z^3 & \dots -141z^5 - 36z^3 & \dots -166z^5 - 36z^3 \\
		&  -24z^5                     &    -25z^5                 &   -32z^5 \\ 
	\end{array}
\end{equation}

\begin{equation}\label{Fig6down}
	\begin{array}{cccc}
		Fig.~\ref{fig:7}  &  A=(1,2);B=(3,4)             &  A=(1,3);B=(2,4)          &  A=(1,4);B=(2,3)   \\
		+origine;down &  \dots - 47z^5 - 16z^3 & \dots -47z^5 -16z^3   &  \dots - 47z^5 - 16 z^3 \\ 
		-A  & \dots - 92z^5  - 24 z^3 & \dots -82z^5  - 24z^3  &  \dots  -93z^5  - 24 z^3 \\
		-B  &  \dots  -77z^5 - 24z^3  &  \dots  -81z^5  - 24z^3 & \dots -84z^5 - 24z^3 \\
		+A,+B & \dots -148z^5 - 36z^3 & \dots -141z^5 - 36z^3 & \dots -160z^5 - 36z^3 \\
		&  -26z^5                     &    -25z^5                 &   -30z^5 \\ 
	\end{array}
\end{equation}

\begin{equation}\label{Fig7up}
	\begin{array}{cccc}
		Fig.~\ref{fig:6}  &  A=(2,3);B=(4,5)             &  A=(2,4);B=(3,5)          &  A=(2,5);B=(3,4)   \\
		+origine;up &  \dots - 31z^5 - 16z^3 & \dots -31z^5 -16z^3   &  \dots - 31z^5 - 16 z^3 \\ 
		-A  & \dots - 69z^5  - 24 z^3 & \dots -52z^5  - 24z^3  &  \dots  -60z^5  - 24 z^3 \\
		-B  &  \dots  -68z^5 - 24z^3  &  \dots  -71z^5  - 24z^3 & \dots -69z^5 - 24z^3 \\
		+A,+B & \dots -136z^5 - 36z^3 & \dots -115z^5 - 36z^3 & \dots -124z^5 - 36z^3 \\
		&  -30z^5                     &    -23z^5                 &   -26z^5 \\ 
	\end{array}
\end{equation}

\begin{equation}\label{Fig7down}
	\begin{array}{cccc}
		Fig.~\ref{fig:7}  &  A=(2,3);B=(4,5)             &  A=(2,4);B=(3,5)          &  A=(2,5);B=(3,4)   \\
		+origine;down &  \dots - 31z^5 - 16z^3 & \dots -31z^5 -16z^3   &  \dots - 31z^5 - 16 z^3 \\ 
		-A  & \dots - 71z^5  - 24 z^3 & \dots -52z^5  - 24z^3  &  \dots  -58z^5  - 24 z^3 \\
		-B  &  \dots  -70z^5 - 24z^3  &  \dots  -71z^5  - 24z^3 & \dots -67z^5 - 24z^3 \\
		+A,+B & \dots -142z^5 - 36z^3 & \dots -115z^5 - 36z^3 & \dots -118z^5 - 36z^3 \\
		&  -32z^5                     &    -23z^5                 &   -24z^5 \\ 
	\end{array}
\end{equation}

The tables~\ref{Fig6up} and~\ref{Fig7up} are associated with the sub-collection $(1,2,3,4)$ of lines; the  line 1 contains two vertical symmetric external fingers in the upper horizontal fragment with magnetic lines. The tables~\ref{Fig6down} and~\ref{Fig7down} are associated with the sub-collection $(2,3,4,5)$ of lines. The  line 5 contains two vertical symmetric internal fingers in the upper horizontal fragment with magnetic lines.  The $\dots$ on tables means hiher terms in the Conway polynomial, which are unrequired to calculate $C_{1,1}$.

The alteration points is taken on the right on the intersection of a vertical "finger" with the horizontal line. The alteration is described by the signed sum of values of one-variable Conway polynomial for corresponding $4$ lines discriminant in each table. For example, the contributions to the alteration  of the bottom row of the first table are defined by two Canway polynomial with $+$ by changing the crossroads of the pair $(A=[1,2];B=[3,4])$,$(A=[1,2];B=[3,4])$,$(A=[1,2];B=[3,4])$ in the corresponding vertical row. The crossroad is the right intersection point on the vertical finger with the horizontal line on the bottom fragment of the diagram Fig.~\ref{fig:6}.

One may see that the internal asymmetry of the skein-relation for the alteration of the each balanced $\Delta$-move, in Figs.~\ref{fig:6} and~\ref{fig:7}, is presented. The signed sum of $4$ coefficients of the term $z^5$ (recall, the considered $4$ coefficients give a recursion of the corresponding $C_{2,2}$-coefficient for $4$-tuple line using the $C_{1,1}$-coefficients of the resolution) by an a corresponding alteration is non-trivial. Of course, jumps of a corresponding $C_{2,2}$-coefficient by a balansed $\Delta$-move are not depend of resolutions of a crossroad.  The sum over the $3$ different partitions of lines in a 4-tuple equals to zero. Coefficients of the term $z^3$ by an alteration alteration are canceled. This means that  the internal asymmetry lost when a cyclic order of lines is destroyed, but the internal asymmetry, itself, is observed.

\section{Applications of the $M$-invariant to the mean field equation with an axion}\label{appl}

In mean magnetic field theory, the construction of exact solutions is abandoned, since such solutions have not yet been generally constructed for a system of MHD equations. Instead, probability distributions for various statistical moments (for example, the energy of the system) of the solution are investigated under the assumption that the solution itself is two-scale and random. When transitioning to an axion field, it can also be assumed that the wave equation of the axion state is described by a random distribution of the pseudo-scalar field $\varphi$, which is a priori stationary relative to a rapidly moving fluid conducting medium with a magnetic field.

It turns out that under this assumption, the mean field equations can take into account the influence of the axion field, which is determined by the new term in the equation, as well as by the fact that the $\alpha$-term is also modified. A diverse range of problems arises; to avoid complicating the equations below, we assume, unless explicitly stated otherwise, that the contribution of high axion frequencies to the statistical ensemble is small and the conductivity of the medium is high, but the limiting transition to infinite conductivity is not studied.

In Refs.~\cite{DvoAkh24,A-D2}, we have derived [with respect to in Eq.~(\ref{alpha})] mean field (dimensionless) equation accounting for the axions contribution:
\begin{equation}\label{meaN}
	\frac{\partial \B}{\partial t} = \frac{1}{\sigma^2_{\rm{c}}} \rot (\b \times \rot \B) + \frac{\dot{\varphi}}{\sigma_{\rm{c}}} \rot \B - \frac{1}{\sigma_{\rm{c}}} \rot \rot \B - \beta \rot \rot \B,
\end{equation}
where $\mathbf{b} = \grad \varphi$ is the axial vector term which is nonzero for inhomogeneous axions, $\varphi = \varphi(\mathbf{x},t)$ is the axion wave-function, $\sigma_c$ is the conductivity, which is taken great $\sigma_{\rm{c}} \to +\infty$, and $\beta$ is the turbulent magnetic diffusion coefficient.  We call this new contribution, $\propto \frac{1}{\sigma^2_{\rm{c}}} \rot (\b \times \rot \B)$ to the induction equation as the axion $\gamma$-term and axion $\alpha$-term, which is $\propto \frac{\dot{\varphi}}{\sigma_{\rm{c}}} \rot \B$, because the third turbulent magnetic diffusion term  $\beta$ is coming from the classical mean MHD.

The axion is a pseudo-scalar field, and we obtain the wave equation (called the Klein-Gordon equation):
\begin{eqnarray}\label{eq7}
	\ddot{\varphi} - \nabla^2\varphi + m^2\varphi = \frac{1}{\sigma_{\rm{c}}}(\rot\B,\B).
\end{eqnarray}
If the conductivity of the medium $\eta_m = \sigma_{\rm{c}}^{-1}$ is not sufficiently large, then equation in Eq.~(\ref{eq7}) is not applicable, since other terms appear on the right-hand side of equation in Eq.~(\ref{eq7}). We assume that the axion mass $m$ is very small and the third term in the left-hand side of the equation is omitted.

Let us prove, using a suitable gauge of time-variable and a scale of the axion field 
$\varphi$, that in the case $\sigma_{\rm{c}} \to +\infty$ the term $\frac{1}{\sigma_{\rm{c}}} \rot \rot \B$ in (\ref{meaN}) is not required. 
The coefficient $\beta$ is not dimensionless, but depends of a spacial scale. The value $\frac{1}{\sigma_{\rm{c}}}-\beta < 0$, when $\beta$ belongs to the turbulent interval of scale. 
In this assumption we get $\frac{1}{\sigma_{\rm{c}}}-\beta \approx -\beta$ and the third term in the right-hand side of (\ref{meaN}) is omitted. In the limit the the approximate inequality becomes exact.

For the simplest case of large $\sigma_{\rm{c}}$, which is introduced in \cite{A-D2}, using the analytic approach of \cite{DvoAkh24}, \cite{AkhDvo24}.  One may see in \ref{eq7}   the first two terms are canceled. This implies that the right-hand side of the equation equals to zero
(magnetic helicity equals to zero) and the  spatial 1D spectrum of the axion field $(\grad \varphi)^2$ ("an anchor") has to be of the greatest order $k^{-1}$, $\grad \varphi \sim k^{-\frac{1}{2}}$. Because the magnetic conductivity $\sigma_{\rm{c}}$ is very large, the $\alpha$-term is blocked (no friction between axion and magnetic modes). This implies that the magnetic field is very small, the spacial scale is described by the Kreichnan spectrum with the magnetic and hydrodynamic energy index $k^{-\frac{3}{2}}$. A balance of the $L^4$-norm of the magnetic energy with the quadratic helicity, the small-scale invariant function in the space of the soliton, is given by the generalized Arnold inequality \cite{A-D2} (Section 3). 

The magnetic conductivities parameter $\sigma_{\rm{c}}$ plays a role of a bifurcation parameter of the system, while  the turbulent conductivity parameter $\beta$ determines spectra of the solution. 
By the equation in Eq.~(\ref{eq7}) the elimination of magnetic field limit $\sigma_{\rm{c}} \to +\infty$ could be compensated by the vanished current helicity term in the right-hand side of the equation. In this case the axion energy becomes much greater then magnetic energy.

Recall, in the  mirror-symmetric simplest state $\sigma_{\rm{c}} \to +\infty$ the solution at each scale consists of a standing and a traveling wave, with the solutions at different scales being independent. The standing waves are due to turbulent diffusion, while the axion itself determines the traveling waves. 
The soliton has zero magnetic helicity.
This state of the soliton is due, firstly, to the absence of the influence of the $\alpha$ term, which generates magnetic modes of the same helicity sign. 
The equation (\ref{helicity4}) has the only second term in the right-hand side. This means that magnetic helicity cannot appear. Because of the Arnold inequality the magnetic energy
can be arbitrary small.
Secondly, to the absence of correlations between the longitudinal $\B_A$ and transverse $\B_B$ magnetic modes of MHD waves of each wave-number \cite{D-A2}.

Assume that the magnetic conductivity $\sigma_{\rm{c}}$ becomes less, but the condition $\frac{1}{\sigma_{\rm{c}}} << \beta$ is satisfied on the large turbulent interval.
The axion is distributed as $\varphi \sim k^{-\frac{5}{3}}$. The first two terms in the left-hand side of the equation (\ref{eq7}) is distributed as $(\B,\rot\B) \sim k^{-\frac{2}{3}}$.
In this case $\grad \varphi \sim k^{-\frac{2}{3}}$,
$\dot{\varphi} \sim k^{-\frac{2}{3}}$. The $\alpha$-effect is observed because of a convolution of two $1D$-distributions in the second $\alpha$-term in the right-hand side of (\ref{meaN}).
This means that the index of $\alpha$-term is $\sim k^{\frac{1}{3}}$. Recall, axion is a pseudo-scalar and the interaction between axion modes and magnetic modes is mirror-asymmetric.
A relation between this interaction with internal asymmetry of magnetic lines is conjectured.
The first term in the right-hand side of (\ref{meaN}) is not helical and gives no contribution to dynamo.
The the third term in the right-hand side of the equation (\ref{meaN}) is still small,  and the first, second and forth terms are distributed identically.

In the left-hand side of Eq.~(\ref{eq7})	the first two terms are distributed as $\sim k^{\frac{1}{3}}$. The difference does not cancel out completely, but only to the first order. The
second-order terms are distributed as $\sim k^{-\frac{2}{3}}$ and the difference equals to the term of the same order in the right-hand side of the equation.

This confirms the statement of Ref.~\cite{A-D2} that an axion mirror soliton by a phase transition transforms into a Kolmogorov soliton, in which the axion energy is partially converted into magnetic energy, and the soliton becomes spiral.
Such a phase transition is likely in an acceleratingly expanding space and is associated with the axion contribution to the $\alpha$ term.

The Kolmogorov MHD spectrum must be left- or right-polarized. In this case, according to Arnold's inequality, the magnetic energy admits a lower bound through magnetic helicity. We recall that, for a mirror soliton, there is no similar lower bound for the magnetic energy, and the magnetic energy of a mirror axion soliton can be arbitrarily small.
This second-order phase transition transforms the mirror magnetic field into a helical one, which is accompanied by an increase in magnetic energy.

An $\alpha$-term appears, which stabilizes across the spectrum.
The Kolmogorov MHD spectrum is observed under the assumption of
polarization of magnetic modes. This follows from the structure of the $M_5$-invariant of magnetic lines, which applies to the mean field magnetic flux.
The $\alpha$-term appears due to the
free energy exchange between axion and magnetic modes. This
occurs for the axion 1D index $\varphi \sim k^{-\frac{5}{3}}$.

We remind that, in Ref.~\cite{A-D2}, we assumed that an initial step of the transition from Kreichnan to Kolmogorov MHD-spectrum with the 1D index
$\B \sim k^{-\frac{5}{6}}$, the main magnetic harmonics, right (left) polarization must occur.
With the assumption that the 1D index of axion $\varphi \sim k^{-\frac{5}{6}}$, it was shown that the characteristic scale at which the $\gamma$-term flux is most intense is determined by the wave number $\propto \eta^{-\frac{2}{3}}$.

By the present more precise consideration, it is naturally to  assume that $\varphi \sim k^{-\frac{5}{3}}$ for the 1D index of axion distribution.
In the 3D distribution, the index of the vector $\b = \nabla \varphi \sim k^{-\frac{8}{3}}$, and the magnetic flux caused by the turbulent diffusion factor is distributed as $\B^{\rm{flow};\beta} \sim k^{-\frac{7}{6}}$, 
see \cite{A9}.
With this assumption  the characteristic scale at which the $\gamma$-term flux is most intense is determined by the wave number $\propto \eta^{-\frac{7}{3}}$.

\section{Conclusion}
Our work is on the border line between solar physics and the dark matter study, which is an unsolved problem in the modern astronomy. We accept the paradigm that dark matter consists of axions. We also mention that one encounters certain difficulties in solar MHD in the numerical and analytical explanation of the presence of intense 3D reconnections without additional 
hypothesis~\cite{R-Y-H-W}. The role of 3D reconnections, which cause nanoflares, in the problem of solar corona heating is described in Ref.~\cite{Kli15}. In our work, we made an attempt to apply axionic MHD for the explanation of this kind of phenomena. Thus, the Sun, as the most studied MHD system, can, to some extent, confirm the existence of axions before these particles are discovered in a laboratory experiment.

We studied the equations for the mean magnetic field in the presence of an axion term, which were defined in \cite{DvoAkh24}.
Two types of fundamental axion MHD spectra were constructed. The properties of these spectra are studied using asymptotic ergodic invariants of magnetic lines, whose densities are gauge-invariant quantities. These quantities determine the indices of the MHD turbulence, since they are transported in the bulk of a conducting magnetic medium by a fast, small scale velocity field as scale-homogeneous quantities.

The simplest magnetic field invariant was defined by Arnold \cite{Arn1}; it is called the asymptotic ergodic Hopf invariant. The type of axion spectrum determined by this invariant coincided with Kraichnan's MHD spectrum; it is characterized by small-amplitude traveling and standing waves generated by the axion.

Another axion MHD spectrum coincides with the  Kolmogorov spectra.
The Kolmogorov spectrum defines a new type of dynamo caused by the presence of an axion. The exponents of all terms in the mean field equation were found to be balanced in the presence of an axion, suggesting that the dynamo solution can be constructed in explicit equations taking into account the axion wave equation. This solution is associated with resonances (including parametric ones) in the system of mean field equations and requires further study.


\begin{thebibliography}{99}
	
	\bibitem{Can19}
	A.~Canepa,
	Searches for supersymmetry at the Large Hadron Collider,
	Rev. Phys. \textbf{4}, 100033 (2019)
	
	\bibitem{Luz20}
	L.~Di~Luzio, M.~Giannotti, E.~Nardi, and L.~Visinelli,
	The landscape of QCD axion models,
	Phys. Rep. \textbf{870}, 1--117 (2020)
	[arXiv:2003.01100].
	
	\bibitem{PecQui77}
	R.~D.~Peccei and H.~R.~Quinn,
	CP Conservation in the Presence of Pseudoparticles,
	Phys. Rev. Lett. \textbf{38}, 1440--1443 (1977).
	
	\bibitem{Wei78}
	S.~Weinberg,
	A New Light Boson?,
	Phys. Rev. Lett. \textbf{40}, 223--226 (1978).
	
	\bibitem{Wil78}
	F.~Wilczek,
	Problem of Strong P and T Invariance in the Presence of Instantons,
	Phys. Rev. Lett. \textbf{40}, 279--282 (1978).
	
	\bibitem{Mar16}
	D.~J.~E.~Marsh,
	Axion Cosmology,
	Phys. Rep. \textbf{643}, 1--79 (2016)
	[arXiv:1510.07633].
	
	\bibitem{Cho21}
	K.~Choi, S.~H.~Im, and C.~S.~Shin,
	Recent progress in the physics of axions and axion-like particles,
	Ann. Rev. Nucl. Part. Sci. \textbf{71}, 225--252 (2021)
	[arXiv:2012.05029].
	
	\bibitem{IraRed18}
	I.~G.~Irastorza and J.~Redondo,
	New experimental approaches in the search for axion-like particles,
	Prog. Part. Nucl. Phys. \textbf{102}, 89--159 (2018)
	[arXiv:1801.08127].
	
	\bibitem{Sik21}
	P.~Sikivie,
	Invisible Axion Search Methods,
	Rev. Mod. Phys. \textbf{93}, 15004 (2021)
	[arXiv:2003.02206].
	
	\bibitem{KimCar10}
	J.~E.~Kim and G.~Carosi,
	Rev. Mod. Phys. \textbf{82}, 557 (2010)
	[arXiv:0807.3125]
	
	\bibitem{Car25}
	P.~Carenza, M.~Giannotti, J.~Isern, A.~Mirizzi, and O.~Straniero,
	Axion Astrophysics,
	Phys. Rept. \textbf{1117}, 1--102 (2025)
	[arXiv:2411.02492]. 
	
	\bibitem{KolTka93}
	E.~W.~Kolb and I.~I.~Tkachev,
	Axion miniclusters and Bose stars,
	Phys. Rev. Lett. \textbf{71}, 3051--3054 (1993)
	[hep-ph/9303313].
	
	\bibitem{KolTka94}
	E.~W.~Kolb and I.~I.~Tkachev,
	Non-Linear Axion Dynamics and Formation of Cosmological
	Pseudo-Solitons,
	Phys. Rev. D \textbf{49}, 5040--5051 (1994)
	[astro-ph/9311037].
	
	\bibitem{ChaDel11}
	P.~H.~Chavanis and L.~Delfini,
	Mass-radius relation of Newtonian self-gravitating
	Bose-Einstein condensates with short-range interactions:
	II. Numerical results,
	Phys. Rev. D \textbf{84}, 043532 (2011)
	[arXiv:1103.2054].
	
	\bibitem{BraZha19}
	E.~Braaten and H.~Zhang,
	Colloquium: The physics of axion stars,
	Rev. Mod. Phys. \textbf{91}, 041002 (2019).
	
	\bibitem{Sik83}
	P.~Sikivie,
	Experimental Tests of the ``Invisible'' Axion,
	Phys. Rev. Lett. \textbf{51}, 1415--1417 (1983);
	Erratum: Phys. Rev. Lett. \textbf{52}, 695 (1984).
	
	\bibitem{Wil87}
	F.~Wilczek,
	Two applications of axion electrodynamics,
	Phys. Rev. Lett. \textbf{58}, 1799--1802 (1987).
	
	\bibitem{LonVac15}
	A.~Long and T.~Vachaspati,
	Implications of a primordial magnetic field for magnetic monopoles, 
	axions, and Dirac neutrinos,
	Phys. Rev. D \textbf{91}, 103522 (2015)
	[arXiv:1504.03319].
	
	\bibitem{DvoSem20}
	M.~Dvornikov and V.~B.~Semikoz,
	Evolution of axions in the presence of primordial magnetic fields,
	Phys. Rev. D \textbf{102}, 123526 (2020)
	[arxiv:2011.12712].
	
	\bibitem{Dvo22}
	M.~Dvornikov,
	Interaction of inhomogeneous axions with magnetic fields
	in the early universe,
	Phys. Lett. B \textbf{829}, 137039 (2022)
	[arxiv:2201.10586].
	
	\bibitem{DilZio03}
	L.~DiLella and K.~Zioutas,
	Observational evidence for gravitationally trapped massive
	axion(-like) particles,
	Astropart. Phys. \textbf{19}, 145--170 (2003)
	[astro-ph/0207073].
	
	\bibitem{Zhi17}
	A.~R.~Zhitnitsky,
	Solar Extreme UV radiation and quark nugget dark matter model,
	J. Cosmol. Astropart. Phys. \textbf{10}, 050 (2017)
	[arXiv:1707.03400].
	
	\bibitem{DvoAkh24}
	M.~S.~Dvornikov and P.~M.~Akhmet'ev,
	Magnetic field evolution in spatially inhomogeneous axion structures, 
	Theor. Math. Phys. \textbf{218}, 515--529 (2024).
	
	\bibitem{AkhDvo24}
	P.~Akhmetiev and M.~Dvornikov,
	Magnetic fields in inhomogeneous axion stars,
	Int. J. Mod. Phys. D \textbf{33}, 2450001 (2024)
	[arxiv:2303.09254].
	
	\bibitem{Dvo24}
	M.~Dvornikov,
	Thin layer axion dynamo,
	Eur. Phys. J. C \textbf{84}, 892 (2024)
	[arxiv:2401.03185].
	
	\bibitem{Dvo25}
	M.~Dvornikov,
	Low mode approximation in the axion magnetohydrodynamics,
	Int. J. Mod. Phys. D, doi:~10.1142/S0218271825410019
	[arxiv:2502.20839].
	
	
	
	\bibitem{Anz23}
	F.~Anzuini, J.~A.~Pons, A.~G\'omez-Ba\~n\'on, P.~D.~Lasky,
	F.~Bianchini, and A.~Melatos,
	Magnetic Dynamo Caused by Axions in Neutron Stars,
	Phys. Rev. Lett. \textbf{130}, 071001 (2023)
	[arXiv:2211.10863].
	
	\bibitem{Sti04}
	M.~Stix,
	\textit{The Sun}
	(Springer, Berlin, 2004), 2nd ed.
	
	\bibitem{Par55}
	E.~N.~Parker,
	Hydromagnetic dynamo models,
	Astrophys. J. \textbf{122}, 293--314 (1955).
	
	\bibitem{Mo}
	H. K. Moffatt, Magnetic Field Generation in Electrically Conducting Fluids, Cambridge Univ. Press, Cambridge (1978).
	
	\bibitem{A-D2} 
	P. M. Akhmetiev, M. S. Dvornikov, {\it Evolution of mirror axion solitons}, Theoret. and Math. Phys., 225:2 (2025), 2017-2029
	
	
	\bibitem{Z-R-S}
	Ya. B. Zeldovich, A. A. Ruzmaikin, and D. D. Sokoloff,  Magnetic Fields in Astrophysics // Gordon and Breach, New York (1983).  
	
	
	\bibitem{Arn1} V. I. Arnold, The asymptotic Hopf invariant and its applications, preprint, Armenian SSR
	Academy of Sciences, Erevan, 1974; Selecta Math. Soviet., 5:4 (1986), 327-345.
	
	
	
	\bibitem{A2}
	P.\,M.~Akhmet'ev, P.M. 
	On a higher integral invariant for closed magnetic lines, revisited
	Journal of Geometry and Physics: 
	170 
	(2021) 
	104379.
	
	
	
	
	\bibitem{K} E. A. Kudryavtseva, {\it Helicity is the Only Invariant of Incompressible Flows whose Derivative is Continuous in the C1 Topology}, Math. Notes, 99:4 (2016), 611-615 math.DS/1511.03746. 
	
	\bibitem{A-Kh} Arnold V. I. and Khesin B. A., Topological Methods in Hydrodynamics Applied Mathematical
	Sciences vol 125 (1998)(2013) (New York: Springer-Verlag).
	
	\bibitem{Ra}
	K.-H. R\"{a}dler, The generation of cosmic magnetic fields, in: From the Sun to the Great Attractor (Guanajuato, August 4-11, 1999, Lecture Notes in Physics, Vol. 556, D. Page and J. G. Hirsh, eds.), Springer, Berlin (2000), pp. 101-172.
	
	
	
	
	
	\bibitem{Arn}
	V. I. Arnold, {\it Arnold's problems}, Springer-Verlag, Berlin, (2004). 
	
	
	\bibitem{A3}	  P. M. Akhmet'ev,  Quadratic helicities and the energy of magnetic fields, Differential equations and dynamical systems, Collected papers, Trudy Mat. Inst. Steklova, 278, MAIK Nauka/Interperiodica, Moscow, 2012, 16-28; Proc. Steklov Inst. Math., 278 (2012), 10-21
	
	
	\bibitem{H}
	P. R. Halmos, \textit{Lectures on Ergodic Theory}, Chelsea, New York (1960).
	
	
	\bibitem{A4}
	P.\,M.~Akhmet'ev
	{\it Topological meaning of the slope of the Kolmogorov spectrum of magnetic turbulence: M5-invariant of magnetic lines and its combinatorial formula}
	J. Geom. Phys.
	178 (2022)
	104583. 
	
	\bibitem{B-M}
	Baader S., March{\'e} J., {\it Asymptotic Vassiliev invariants for vector fields}, Bull. Soc.
	math. France, 140:4 (2012), 569-582
	
	
	\bibitem{S} Sokoloff D.D., {\it Mirror Asymmetry and Helicity Invariants in Astrophysical Dynamos}, Geomagn. Aeron., 59:7 (2019), 799-805
	
	\bibitem{D}
	P. Dehornoy, Asymptotic invariants of 3-dimensional vector fields, Winter Braids Lecture Notes, 2 (2016), 1
	
	
	\bibitem{A-V}
	P. M. Akhmet'ev, I. V. Vyugin, {\it Dispersion of the Arnold's asymptotic ergodic Hopf invariant and a formula for its calculation}, Arnold Math. J., 2020, published 18.06.2020, https://doi.org/10.1007/s40598-020-00144-w; arXiv: 1906.12131
	
	
	
	\bibitem{P-S}
	V.\,V.~Prasolov, A.\,B.~Sossinsky
	Knots, Links, Braids and 3-
	Manifolds
	Trans. Math. Monogr.,
	Amer. Math. Soc., Providence RI,
	1997
	154.
	
	
	\bibitem{A-S}
	P.M.Akhmetiev, S.Yu.Smirnov {\it
		Combinatorial formula for the $M$-invariant of magnetic lines I,II} Differential equations and related mathematical qurestions, XVII Conference, Kolomna, Russia (2025).
	
	
	\bibitem{A5}
	P.\,M.~Akhmetiev 
	{\it On a new integral formula for an invariant of 3-component oriented links},
	J. Geom. Phys.
	53 (2), 180-196 (2005).
	
	
	
	
	
	
	
	
	
	
	
	\bibitem{A7}	
	P. M. Akhmet'ev, {\it Knot Invariants in Geodesic Flows}, Differential equations and dynamical systems,  Proc. Steklov Inst. Math., 308 (2020), 42-55.
	
	
	\bibitem{A-D} 
	P.M.Akhmet'ev and M.S.Dvornikov,  Invariants of magnetic lines for Yang-Mills solutions, 
	Journal of Geometry and Physics,  https://doi.org/10.1016/j.geomphys.2024.105102 (2024).
	
	
	
	\bibitem{D-A2} 
	M. S. Dvornikov and P. M. Akhmetiev, {\it Evolution of a Large Scale Chern-Simons Magnetic Field in Expanding Space with a Negative Scalar Curvature Parameter}, Physics
	of Particles and Nuclei, 2025, Vol. 56, No. 2, pp. 269-274.
	
	\bibitem{R-Y-H-W} 
	A.J.B. Russell, A.R. Yeates, G. Hornig, A.L. Wilmot-Smith, Evolution of field line helicity during magnetic reconnection, Phys. Plasmas 22 (2015) 032106, 
	arXiv:1501.04856v2[physics .plasm -ph], 2015.
	
	
	
	
	
	\bibitem{Kli15}
	J.~A.~Klimchuk,
	Key Aspects of Coronal Heating,
	Phil. Trans. R. Soc. A: Math. Phys. Eng. Sci. \textbf{373}, 20140256 (2015)
	[arXiv:1410.5660].
	
	
	\bibitem{A8}
	P.M. Akhmetyev, {\it On Arnold's problem on a higher analogue of the Hopf asymptotic invariant},
	J Math Sci 208, 24-35 (2015). https://doi.org/10.1007/s10958-015-2420-7
	
	
	\bibitem{A-M-R} P. M. Akhmet'ev, D. Repov$\breve{s}$, I. Maleshich, {\it On Milnor's Invariants of 4-Component Links}, Mat. Zametki, 71:4 (2002), 495-507; Math. Notes, 71:4 (2002), 455-463 
	
	\bibitem{A9} P. M. Akhmet'ev, {\it Topological meaning of the slope of the Kolmogorov spectrum of magnetic turbulence}, TMF, 209:2 (2021), 351-366; 
\end{thebibliography}
\end{document}